# Rietveld Refinement and NMR Crystallographic Investigations of Multicomponent Crystals Containing Alkali Metal Chlorides and Urea

Cameron S. Vojvodin,[1,2] Sean T. Holmes,[1,2] Christine E. A. Kirschhock,[3] David A. Hirsh,[4] Igor Huskić,[5] Sanjaya Senanayake,[6] Luis Betancourt,[6] Wenqian Xu,[7] Eric Breynaert,[3,8] Tomislav Friščić,[5,9] and Robert W. Schurko[1,2]*

[1] Department of Chemistry & Biochemistry, Florida State University, Tallahassee, FL, 32306, USA
[2] National High Magnetic Field Laboratory, Tallahassee, FL, 32310, USA
[3] Centre for Surface Chemistry and Catalysis – Characterisation and Application Team, KU Leuven, 3001 Leuven, Belgium
[4] Department of Chemistry & Biochemistry, University of Windsor, Windsor, Ontario N9B 3P4, Canada
[5] McGill University, Montreal, Quebec H3A 0B8, Canada
[6] Chemistry Division, Brookhaven National Laboratory, Upton, New York 11973, USA
[7] X-ray Science Division, Advanced Photon Source, Argonne National Laboratory, Argonne, Illinois 60439, USA
[8] NMR/X-ray platform for Convergence Research (NMRCoRe), KU Leuven, 3001 Leuven, Belgium
[9] School of Chemistry, University of Birmingham, Edgbaston, Birmingham, B15 2TT, UK

*Author to whom correspondence should be addressed.

E-mail: rschurko@fsu.edu

**Abstract**

New mechanochemical preparations of three multicomponent crystals (MCCs) of the form MCl:urea·$nH_2O$ (M = $Li^+$, $Na^+$, and $Cs^+$) are reported. Their structures were determined by an NMR crystallography approach, combining Rietveld refinement of synchrotron powder X-ray diffraction data (PXRD), multinuclear ($^{35}Cl$, $^{7}Li$, $^{23}Na$, and $^{133}Cs$) solid-state NMR (SSNMR) spectroscopy, and thermal analysis. Mechanochemical syntheses of the three MCCs, two of which are novel, are optimized for maximum yield and efficiency. $^{35}Cl$ SSNMR is well-suited for the structural characterization of these MCCs since it is sensitive to subtle differences and/or changes in chloride ion environments, providing a powerful means of examining H···$Cl^-$ bonding environments. Alkali metal NMR is beneficial for identifying the number of unique magnetically and crystallographically distinct sites and enables facile detection of educts and/or impurities. In the case of NaCl:urea·$H_2O$, $^{23}Na$ MAS NMR spectra are key, both for identifying residual NaCl educt and for monitoring NaCl:urea·$H_2O$ degradation, which appears to proceed via an autocatalytic decomposition process driven by water (with a rate constant of $k = 1.22 \times 10^{-3}$ s$^{-1}$). SSNMR and PXRD were used to inform the initial structural models. Following Rietveld refinement, the models were subjected to dispersion-corrected plane-wave DFT geometry optimizations and subsequent calculations of the $^{35}Cl$ EFG tensors, which enable the refinement of hydrogen atom positions, as well as the exploration of their relationships to the local hydrogen bonding environments of the chloride ions and crystallographic symmetry elements.

## 1. Introduction

An active area of research in crystal engineering is the synthesis, characterization, and rational design of multicomponent crystals (MCCs), including cocrystals, salts, hydrates, and their permutations.[1–5] Pharmaceutical cocrystals are an important class of MCCs, consisting of at least one or more active pharmaceutical ingredients (API) and one or more pharmaceutically-acceptable coformers.[6,7] Pharmaceutical cocrystals are increasingly important in the pharmaceutical industry since it is possible to rationally design products exhibiting specific physicochemical properties (*i.e.*, stability, solubility, bioavailability, and tableting behavior) with careful consideration of coformers and synthetic methods.[8–11] Ionic cocrystals involving metal halides (*e.g.*, LiCl, NaCl, KCl, *etc.*) are excellent candidates for designing new solid forms of neutral APIs (*i.e.*, they might not contain ionizable functional groups) since they offer additional ways to enhance and tailor physicochemical properties. Consequently, there is a need for reaction strategies and techniques for rapidly generating high yields of phase-pure MCCs and/or pharmaceutical cocrystals comprising diverse constituents, alongside reliable methods for their characterization.

*Mechanochemistry* encompasses a class of techniques that employs mechanical actions (*i.e.*, rubbing, grinding, or shearing) to induce chemical and/or physical transformations, typically involving solid reactants.[12–14] In recent years, mechanochemistry has established itself as an important synthetic tool for the production of a wide variety of materials, including organic crystals, metal-organic frameworks, and pharmaceuticals.[12,13,15] A common mechanochemical method is ball milling, where solid reactants (or educts) and ball bearings are placed in a sealed milling jar, which is agitated using a mechanical vibration or rotation stage to induce a reaction. Ball milling can involve neat grinding (NG), where the dry solid reactants are ground together, and liquid-assisted grinding (LAG),[16] where a small amount of liquid (often μL quantities) is added to increase reaction efficiency, act as a catalyst, and/or prepare novel solid forms unobtainable by conventional synthetic pathways.[11,13,14,17–20] Mechanochemistry has garnered much interest due to its adherence to the tenets of *green chemistry*, including minimal solvent use, low-energy input, high atom economy, and reduced waste and/or by-products.[13,17,20–22]

The characterization of solid forms emerging from mechanochemical reactions is essential for identifying and determining structures of products, by-products, and/or the presence of residual educts. Since these solids are often microcrystalline powders, single-crystal XRD (SCXRD) is not feasible. There are many alternative techniques for their characterization, including powder XRD (PXRD), thermal gravimetric analysis (TGA), differential scanning calorimetry (DSC), vibrational spectroscopy (IR and Raman), and solid-state NMR (SSNMR) spectroscopy.[10,23,24] Notably, if high-quality PXRD data can be acquired, it is possible to determine crystal structures via Rietveld refinement. However, there are numerous challenges associated with Rietveld refinement, including sample quality, data quality (*i.e.*, S/N ratios, limited angular ranges, systematic errors), peak overlap from multiple phases, the selection of an appropriate refinement model (*i.e.*, space group, atomic positions, *etc*.), the choice of refinement constraints for convergence, and the need for a proficient level of expertise in crystallography.[25–27] Since SSNMR can give detailed information on the molecular-level structure, local environments, and even interatomic proximity, it is well-positioned as a complementary technique to aid in the crystallographic characterization of products of mechanochemical reactions.

*NMR crystallography* (NMRX) is a broad term describing the combined use of SSNMR spectroscopy, XRD methods, and quantum chemical calculations to solve, validate, and/or refine, crystal structures.[28–35] Most modern NMRX studies rely on comparisons of experimentally-measured chemical shifts (commonly $^{1}H$, $^{13}C$, and $^{15}N$) to those obtained from theory to both guide the solution of crystal structures and assess crystal structure quality.[36–48] Although chemical shifts can be measured with relative ease, they are often challenging and time-consuming to calculate accurately from first principles, making them of limited use for NMRX studies in certain cases (*e.g.*, chemical shifts tensors that are only marginally impacted by their longer range environment).

Measurements and first-principles calculations of quadrupolar interaction (QI) parameters can provide an alternative route for NMRX.[49–59] SSNMR spectra of quadrupolar ($I > 1/2$) nuclei almost always feature powder patterns that are strongly influenced by the QI. With careful experimentation, it is possible to measure not only the isotropic chemical shift ($\delta_{iso}$), but the quadrupolar coupling constant ($C_Q$) and asymmetry parameter ($\eta_Q$), which can be compared to electric field gradient (EFG) tensors derived from computations (see **Table 1** for definitions). EFG tensors are exquisitely sensitive to subtle changes or differences in solid-state structures, including weaker, long-range interactions (*e.g.*, hydrogen bonding, π-π stacking, and halogen bonding), thereby providing a source of structural information distinct from chemical shifts in both origin and scope.[42,43,55,60–64] Moreover, DFT calculations of EFG tensors are much less computationally expensive than those of chemical shift tensors, which could aid in accelerating NMRX routines. As such, it is desirable to develop and employ quadrupolar NMR crystallography (QNMRX) for structure validation and refinement, NMRX-guided Rietveld refinements,[65–67] and NMRX-assisted crystal structure prediction.[40,51,52,54,59,62,68–71]

Of the quadrupolar nuclides commonly found in organic solids, $^{35}Cl$ ($I = 3/2$) is among the most widely investigated, due to the relative ease of $^{35}Cl$ SSNMR experiments and prevalence of HCl salts of organic solids.[58,72–86] $^{35}$Cl SSNMR spectra typically feature broad central transition (CT, +1/2 ↔ –1/2) powder patterns that are influenced by the second-order quadrupolar interaction (SOQI) and chemical shift anisotropy (CSA). Notably, the $^{35}$Cl EFG tensors of chloride ions are extremely sensitive to their local hydrogen bonding environments, namely, the number of H···Cl$^-$ hydrogen bonds, their H···Cl$^-$ distances, and their spatial arrangements.[55–58] Therefore, $^{35}$Cl SSNMR spectra provide unique spectral fingerprints for each magnetically and crystallographically distinct chloride ion, enabling straightforward differentiation of polymorphs, hydrates, and other solid forms, including MCCs.[54,57,58,77,82–84]

Herein, we discuss the mechanochemical preparation of three urea-containing MCCs of the form MCl:urea·$n$H$_2$O (M = Li, Na, and Cs; $x$ = 0, 1). Urea is an ideal candidate molecule for developing multinuclear QNMRX methods since it is a small and simple molecule capable of engaging in hydrogen bonding as both a donor and an acceptor, which has relevance in drug design and delivery for establishing drug-target interactions. We also present the QNMRX characterization using synchrotron PXRD, thermal analysis, and multinuclear SSNMR, in tandem with Rietveld refinements and plane-wave DFT calculations. Novel mechanochemical syntheses of the three MCCs were optimized for maximum yield and efficiency. The identities and purities of all MCCs and educts were confirmed with PXRD and multinuclear SSNMR; in the case of NaCl:urea·H$_2$O, PXRD data were compared to simulated patterns derived from the previously-reported SCXRD structure.[87] To aid the Rietveld refinements, all data were used in concert to inform the initial structural inputs. The final structural solutions were then subjected to dispersion-corrected plane-wave DFT geometry optimizations to identify the structural models with the lowest static lattice energies. Lastly, DFT calculations of the $^{35}$Cl EFG tensors were conducted on the geometry-optimized structures to assess the agreement between experiment and theory, and to elucidate relationships between NMR parameters and molecular-level structures.

## 2. Experimental

### 2.1 Materials

Alkali metal chlorides (MCl, M = Li, Na, K, Rb, and Cs), urea, and solvents were purchased from Sigma Aldrich Ltd. Solid reagents were dried *in vacuo* (120 $^{o}$C, 15 mm Hg) for a minimum of 16 h before use.

### 2.2 Mechanochemical Synthesis of Multi-Component Crystals

CsCl:urea and LiCl:urea were prepared mechanochemically via NG, whereas NaCl:urea·$H_2O$ was prepared through LAG with 10 μL of triethylamine (TEA) used as the milling liquid additive ($\eta$ = 0.236 μl/mg) (**Table S1**). Syntheses were conducted using a Retsch Mixer Mill 400, 10 mL stainless steel milling jars, two 7 mm stainless steel ball bearings (ball weight: 1.384 g), a milling frequency of 30 Hz and, in most cases, a total milling time of up to 40 minutes. Unless indicated otherwise, syntheses were scaled to a total weight of *ca.* 200 mg for the solid reagents to ensure that ball milling conditions are as similar as possible between different reactions.

### 2.3 Thermogravimetric Analysis (TGA) and Differential Scanning Calorimetry (DSC)

Simultaneous TGA and DSC measurements were performed on a TA SDT Q600 instrument using an alumina crucible. Samples were heated from room temperature (23 ˚C) to 500 °C at a heating rate of 10 °C min$^{-1}$ under a dry argon purge (gas flow of 100 mL min$^{-1}$). Approximately 5 mg of sample was used for all measurements.

## 2.4 Powder X-Ray Diffraction

The PXRD patterns were acquired using a Proto AXRD benchtop diffractometer with a Cu $K_{\alpha}$ ($\lambda$ = 1.540593 Å) radiation source and a Proto DECTRIS hybrid pixel detector. NaCl:urea·$H_2O$ and CsCl:urea were mounted on a Proto plate sample holder, whereas LiCl:urea was mounted on an air-sensitive sample holder. All diffraction experiments were conducted with an X-ray tube voltage of 30 kV, a current of 20 mA, 2θ angles ranging from 5 - 50°, a step size of 0.015°, and a dwell time of 5 s (resulting in an acquisition time of *ca*. 35 minutes per sample). PXRD patterns were processed and simulated using the CrystalDiffract software package.

## 2.5 Synchrotron X-ray Diffraction

High-resolution synchrotron PXRD patterns were collected at the ANL-APS beamline 17-BM at the Advanced Photon Source (APS) at Argonne National Laboratory (ANL). A NIST standard of lanthanum hexaboride ($LaB_6$) powder was used to calibrate the sample-to-detector distance (400, 700, and 1000 mm) and the synchrotron X-ray wavelength ($\lambda$ = 0.45390 Å), and a Si flat detector (PerkinElmer) was used to collect two-dimensional XRD patterns. Samples were loaded into a 1 mm o.d. Kapton tubes. All measurements were conducted at 295 K.

## 2.6 Solid-State NMR Spectroscopy

**2.6.1 Overview.** All moderate-field NMR [$B_0$ = 9.4 T, $\nu_0(^7Li)$ = 155.51 MHz, $\nu_0(^{23}Na)$ = 105.84 MHz, $\nu_0(^{133}Cs)$ = 52.48 MHz, $\nu_0(^{35}Cl)$ = 39.21 MHz] experiments were conducted on a Bruker Avance III HD NMR spectrometer with an Oxford 9.4 T wide-bore magnet at the University of Windsor (Windsor, Canada). Static experiments were conducted using a Varian/Chemagnetics 4 mm triple-resonance HXY magic-angle spinning (MAS) probe, whereas MAS NMR experiments were conducted using a Varian/Chemagnetics 4 mm double-resonance HX MAS probe. Samples were packed in air-tight 4 mm o.d. zirconia rotors for both sets of experiments. MAS experiments were stabilized at 5 ˚C (278 K) using a Varian variable-temperature upper stack and nitrogen gas heat exchanger to prevent sample decomposition due to frictional heating in the rotor. High-field NMR [$B_0$ = 21.1 T, $\nu_0(^{35}Cl)$ = 88.14 MHz, $\nu_0(^1H)$ = 900.00 MHz] experiments were conducted on a Bruker Avance II NMR spectrometer equipped with an Oxford 21.1 T standard-bore magnet at the National Ultrahigh Field NMR Facility for Solids (Ottawa, Canada). Static experiments were acquired using a static home-built 4 mm double-resonance HX low-E probe, whereas MAS NMR experiments were conducted using a 4 mm double-resonance HX MAS Bruker probe. All samples were packed in 4 mm o.d. zirconia rotors. MAS samples were cooled to 5 ˚C using a Bruker BVT-3000 and dry nitrogen gas to prevent MAS heating and subsequent sample decomposition. Detailed acquisition parameters for all SSNMR experiments are found in the ESI (**Table S2**). All data was processed in Bruker TopSpin version 4.1.1,[88] and spectral fitting was carried out using the ssNake software package.[89] To ensure proper expression of the Euler angles (describing the relative orientation of the EFG and CS tensors) in the *ZY′Z″* convention,[90–92] the results of simulations in ssNake, which uses the *ZX′Z″* convention and different definitions for anisotropic chemical shift and quadrupolar parameters, were verified in WSOLIDS1.[93] Euler angles were converted from the *ZY′Z″* convention for direct comparison to tensor orientations extracted from CASTEP calculations (*vide infra*) using the EFGShield software package,[90] which uses the *ZY′Z″* convention.[94] Uncertainties were assessed through bidirectional variation of each parameter *via* comparison of experimental and simulated spectra.

**2.6.2 $^{7}$Li SSNMR Spectroscopy.** $^{7}$Li MAS NMR spectra were acquired at 9.4 T using a rotor-synchronized Hahn-echo sequence ($\nu_{rot}$ = 5 kHz) and high power $^{1}$H decoupling ($\nu_2$ = 25 kHz), with a 2.5 μs non-selective π/2 pulse ($\nu_1$ = 100 kHz), an optimized recycle delay of 10 s, and 1024 scans. Chemical shifts were referenced to 1.0 M LiCl(aq) in $D_2O$ ($\delta_{iso}(^{7}Li)$ = 0.0 ppm).

**2.6.3 $^{23}$Na SSNMR Spectroscopy.** $^{23}$Na MAS NMR spectra were acquired at 9.4 T with Bloch decay and Hahn-echo pulse sequences, with $\nu_{rot}$ = 10 kHz, high-power $^{1}$H decoupling ($\nu_2$ = 50 kHz), 2.5 μs CT-selective π/2 pulses ($\nu_1$ = 50 kHz), an optimized recycle delay of 25 s, and 32 scans. Chemical shifts were referenced to 0.1 M NaCl (aq) in $D_2O$ ($\delta_{iso}(^{23}Na)$ = 0.0 ppm). To monitor the decomposition of NaCl:urea·$H_2O$ MCCs, variable-temperature $^{23}$Na MAS NMR spectra were collected with a recycle delay of 37.5 s and 16 scans, resulting in a total experiment time of 10 min. Integrated intensities of the powder patterns in the $^{23}$Na MAS NMR spectra were used to monitor the decomposition of the MCC into the starting educts.

**2.6.4 $^{133}$Cs SSNMR Spectroscopy.** $^{133}$Cs MAS NMR spectra were acquired at 9.4 T with a rotor-synchronized Hahn-echo pulse sequence with a spinning speed of $\nu_{rot}$ = 5 kHz, no $^{1}$H decoupling, 4.0 μs non-selective π/2 pulses ($\nu_1$ = 62.5 kHz), an optimized recycle delay of 90 s, and 32 scans. Chemical shifts were referenced to 1.0 M CsCl (aq) in $D_2O$ ($\delta_{iso}(^{133}Cs)$ = 0.0 ppm).

**2.6.5 $^{35}$Cl SSNMR Spectroscopy.** The $^{35}$Cl NMR spectra were acquired at $B_0$ = 9.4 T and 21.1 T under static and MAS conditions. At 9.4 T, static and MAS ($\nu_{rot}$ = 12 kHz) spectra were acquired with a Hahn-echo sequence with high-power $^{1}$H decoupling ($\nu_2(^{1}H)$ = 25 kHz), and at 21.1 T, the static spectra were acquired with a quadrupolar echo sequence (90°-τ-90°-acq) with high-power $^1H$ decoupling ($\nu_2(^1H)$ = 35 kHz). MAS spectra were acquired with a Bloch decay pulse sequence and $^1H$ decoupling ($\nu_2(^1H)$ = 25 kHz). Chemical shifts were referenced to solid NaCl ($\delta_{iso}(^{35}Cl)$ = 0.0 ppm).

**2.6.6 $^{35}$Cl Variable-Temperature (VT) NMR.** $^{35}$Cl VT-NMR experiments were conducted on a Bruker NEO spectrometer with an Oxford 18.8 T narrow-bore magnet ($\nu_0(^{35}Cl)$ = 78.383 MHz) at the National High Magnetic Field Laboratory (NHMFL) in Tallahassee, Florida. Spectra were acquired with a home-built 5mm HX static probe under static conditions (*i.e.*, stationary samples) with samples packed into 5 mm polychlorotrifluoroethylene sample holders with Viton o-rings designed at the NHMFL and machined by Shenzhen Rapid Direct Co., Ltd. Spectra were acquired with a QCPMG pulse sequence employing CT-selective pulses, calibrated recycle delays at all temperatures, and a continuous-wave (CW) $^1H$ decoupling field of 40 kHz. See **Table S3** for further details. $^{35}$Cl chemical shifts were referenced with respect to 0.1 M NaCl (aq) at $\delta_{iso}(^{35}Cl)$ = 0 ppm using NaCl(s) as a secondary reference at $\delta_{iso}(^{35}Cl)$ = –41.11 ppm.

## 2.7 Density-Functional Theory (DFT) Calculations

**2.7.1 Overview.** All quantum chemical calculations were performed within the framework of plane-wave DFT as implemented in the CASTEP module of BIOVIA Materials Studio 2020,[95] with structural models derived from the Rietveld refinements presented herein (see **§3.3** for details), or single-crystal X-ray structures of LiCl, NaCl, and CsCl.[96–98] All DFT calculations employed the RPBE functional,[99] ZORA ultrasoft pseudopotentials generated-on-the-fly,[100] and a version of the Tkatchenko and Scheffler dispersion correction force field (DFT-TS),[101] which was reparametrized (*i.e.*, DFT-TS*) to aid in refining crystal structures of organic solids.[40] The TS* semiempirical dispersion correction was selected for this work because it is parameterized for elements up to and including the sixth period.

**2.7.2 Geometry Optimizations.** Geometry optimization used the LBFGS energy minimization scheme,[102] while holding the unit cell parameters constant. Calculations used an SCF convergence threshold of $5 \times 10^{-6}$ eV atom$^{-1}$, a plane-wave cut-off energy of 800 eV, and evaluated integrals over a Brillouin zone using a Monkhorst-Pack grid with *k*-point spacing of 0.05 Å$^{-1}$.[103] Structural convergence was determined using a maximum change in energy of $5 \times 10^{-6}$ eV atom$^{-1}$, a maximum displacement of $5 \times 10^{-4}$ Å atom$^{-1}$, and a maximum Cartesian force of $10^{-2}$ eV Å$^{-1}$.

**2.7.3 NMR Interaction Tensors.** $^{35}$Cl EFG tensors were calculated from structural models obtained from DFT-TS* geometry optimizations (*vide infra*). Magnetic shielding tensors were calculated with the gauge including projector augmented wave (GIPAW) approach.[104–106] Conversion between the magnetic shielding and chemical shift scales was accomplished through the following procedures: (i) $^{35}$Cl shifts were referenced to NaCl(s) [$\delta(^{35}$Cl) = 0.0 ppm; $\sigma(^{35}$Cl) = 995.7 ppm]; (ii) $^{7}$Li shifts were referenced to 1.0 M LiCl(aq) ($\delta(^{7}$Li) = 0.0 ppm) using calculations on LiCl(s); [$\delta(^{7}$Li) = –1.1 ppm, $\sigma(^{7}$Li) = 90.8 ppm]; (iii) $^{23}$Na shifts were referenced to 0.1 M NaCl(aq) ($\delta(^{23}$Na) = 0.0 ppm) using calculations on NaCl(s) [$\delta(^{23}$Na) = 7.0 ppm; $\sigma(^{23}$Na) = 553.0 ppm]; (iv) $^{133}$Cs shifts were referenced to 1.0 M CsCl(aq) ($\delta(^{133}$Cs) = 0.0 ppm) using calculations on CsCl(s) [$\delta(^{133}$Cs) = 223.2 ppm; $\sigma(^{133}$Cs) = 5650.4 ppm]. Euler angles describing the relative orientations of the EFG and CS tensors were extracted from the CASTEP output files using EFGShield 4.1,[90] and correspond to the *ZY′Z″* convention.[90–92]

## 2.8 Molecular Mechanics Calculations

Molecular mechanics geometry optimizations were performed using the FORCITE module within Materials Studio 2020. These structural refinements employed the COMPASS III force field[107,108] and an atom-based summation method for all interaction energies (*i.e.*, electrostatic, van der Waals, and hydrogen bonding). The molecular mechanics geometry optimizations were performed using the following convergence thresholds: $2 \times 10^{-6}$ kcal mol$^{-1}$ for energy, $10^{-3}$ kcal mol$^{-1}$ Å$^{-1}$ for forces, and $10^{-6}$ Å for structural displacement.

## 2.9 Initial Structure Identification

Synchrotron PXRD patterns were indexed using the X-Cell algorithm, as implemented in the REFLEX Powder Indexing module of Materials Studio 2020,[109] to determine the unit cell parameters. Subsequent Pawley refinement assisted to identify the most probable space group.[110] Unit cells were then populated with the constituent atoms in the appropriate stoichiometric ratios (according to the stoichiometries determined by a combination of multinuclear SSNMR, indexing of PXRD, TGA/DSC, and chemical intuition based on reaction stoichiometries) and subjected to a molecular mechanics calculation using FORCITE. The structures of these MCCs were subjected to a series of alternating REFLEX refinements and CASTEP geometry optimizations, until a single structural model consistent with both methods was achieved. The CASTEP geometry-optimized structural model was then submitted for Rietveld refinement.

## 2.10 Structure Solution via Rietveld Refinement

All structural solutions and Rietveld refinements were conducted using the synchrotron PXRD data. The resulting structural models were optimised in FOX.[111] Suggested space groups were verified or adjusted as necessary. The space group with highest symmetry that accurately describes the PXRD data was chosen throughout. Non-standard settings of monoclinic space groups were chosen when their monoclinic angles were found to be close to 90°. The optimized structures served as starting model for GSAS.[112] Initially, urea was refined as rigid body; however, towards the end of the refinement, all molecules were freely refined. In the case of NaCl:urea·$H_2O$, NaCl was added as a second phase in the refinement, and was determined as 0.6% of the total weight (see **§3.2.2** for further details).

## 3. Results and Discussion

**3.1 Mechanochemical Syntheses.** Mechanochemical syntheses (NG and LAG) using alkali metal chlorides (MCl, M = Li, Na, K, Rb, and Cs) and urea as starting reagents were attempted, but only three MCCs were successfully produced: NaCl:urea·$H_2O$, CsCl:urea, and LiCl:urea (see **Table S1** for a summary of ball milling experiments), which were confirmed with PXRD (**Figure 1**; also see **Figure S1** for PXRD patterns for solid products of unsuccessful syntheses). NaCl:urea·$H_2O$, which was previously generated by slow evaporation from $H_2O$,[87] was here prepared with LAG using 10 μL of TEA. Comparison of the experimental and simulated PXRD patterns indicates a small amount of unreacted NaCl (determined by $^{23}$Na SSNMR to be *ca*. 5 % w/w, *vide infra*). By contrast, the novel CsCl:urea and LiCl:urea MCCs were prepared mechanochemically *via* NG, with their PXRD patterns showing no evidence of educts or impurity phases. Initial trials revealed that all three MCCs could be prepared in 40 min, a significant improvement over the growing crystals from solution, which can take several days to weeks. Remarkably, NaCl:urea·$H_2O$ and CsCl:urea can be made significantly faster with LAG using TEA, with optimized syntheses taking only five minutes to yield pure products (**Figure S2** and **Figure S3**). The mechanochemical synthesis of LiCl:urea was not optimized due to sample deliquescence, with shorter milling times resulting in residual educts detectable by PXRD.

Despite their simple composition, these three MCCs proved challenging to prepare. The formation of NaCl:urea·$H_2O$ was found to be highly dependent upon the pH of the water. Syntheses involving deionized (DI) water, which has a pH ≈ 5.0 – 5.5 due to the dissolution of $CO_2$(g), always resulted in significant excesses of NaCl; we posit that this is likely due to carbonic acid in the DI water that deters MCC formation. To confirm this, four ball milling trials, conducted with (i) 0.1 M $HNO_3$ in DI water (pH ≈ 1), (ii) neat DI water (pH ≈ 5.5), (iii) Kirkland Signature™ bottled water (pH ≈ 6), and (iv) 10 μL TEA in 18 μL DI water (pH ≈ 11), show that with increasing pH, less NaCl educt is present in the ball milled products (**Figure S4**). However, the use of bottled water with minimal atmospheric exposure results in the fewest impurities. Thus indicating that the pH and alkalinity of water of the water used for synthesis is crucial in preparing products with high purity. The TGA/DSC scan of NaCl:urea·$H_2O$ indicates MCC decomposition with an onset temperature of *ca*. 50 °C, followed by loss of water at *ca*. 110 °C (**Figure S5A**). As such, DSC indicates that NaCl:urea·$H_2O$ is unstable under ambient conditions, whereas the TGA confirms that it is a monohydrate. The instability of NaCl:urea·$H_2O$ was further explored using $^{23}$Na NMR (*vide infra*).

Since CsCl:urea was assumed to be a hydrate, its initial preparations occurred similarly to NaCl:urea·$H_2O$, using LAG with one equivalent of $H_2O$. The resulting PXRD pattern for the LAG product **(Figure S6**) indicated the formation of a novel MCC and no leftover CsCl. However, TGA/DSC analysis (**Figure S5B**) showed no evidence of water loss prior to the decomposition of the MCC at *ca*. 200 °C. As this indicated CsCl:urea to be anhydrous, this compound was also prepared *via* NG (without water). PXRD confirmed the NG product to be the same phase (**Figure S7**).

Preparation of LiCl:urea involves reagents and a final product that are both hygroscopic. The mixture resulting from ball milling appeared as a grey, wet paste. PXRD of the product revealed the presence of a significant amount of unreacted LiCl. As water was suspected to be the culprit, in a next attempt, reagents were filled into the milling jar in a drybox ($N_2$ atmosphere, <0.1% RH). The milling jar was wrapped with Teflon tape to exclude air and moisture during subsequent ball milling in ambient conditions. PXRD revealed the resulting product to be a novel MCC, without impurities. TGA/DSC (**Figure S5C**) revealed minor losses of water at *ca*. 50 and 130 ˚C, prior to decomposition of the product at *ca*. 240 ˚C. Because these losses were small in comparison to those observed for NaCl:urea·$H_2O$, they are attributed to the removal of surface water. This suggests LiCl:urea also is anhydrous.

We note that our failure to produce MCCs involving $K^+$ and $Rb^+$ is unsurprising in light of the previous work by Braga and co-workers,[113] where they propose that the ratio between the ionic radii of the cation and anions plays a role in cocrystallization and dictates the type of crystal packing. They observe that that the $K^+/Cl^-$ ionic radius ratio is too large to form a 1:1 MCC and too small for form a 1:2 MCC.  This is likely true for the case of $Rb^+$ as well.

## 3.2 Multinuclear SSNMR Spectroscopy

**3.2.1 $^{35}$Cl SSNMR Spectroscopy.** $^{35}$Cl SSNMR spectra (**Figure 2**) were acquired at two magnetic fields (9.4 T and 21.1 T) to aid in the precise measurement of the EFG and CS tensors, since the broadening of CT patterns scales with $B_0^{-1}$ and $B_0$ and for the SOQI and CSA, respectively.[114–116] $^{35}$Cl MAS NMR spectra allow for accurate determination of $C_Q$, $\eta_Q$, and the isotropic chemical shift, $\delta_{iso}$, since the effects of CSA on the relatively narrow CT patterns are completely or partially averaged at high enough MAS rates ($\nu_{rot}$ = 10 – 12 kHz herein), leaving only the partially averaged effects of the SOQI. These parameters aid in fitting the static CT powder patterns, along with the span (Ω) and skew (κ) of the CS tensor, and the Euler angles defining the relative orientation of the EFG and CS tensors (α, β, and γ) (see **Table 1** for definitions of these parameters).

In all cases, the $^{35}$Cl NMR spectra feature single broad CT patterns indicating no residual educts or impurity phases, with the exception of the MAS NMR spectrum of NaCl:urea·$H_2O$ acquired at 9.4 T, where a sharp peak is observed at *ca.* –41.1 ppm, corresponding to a small amount of unreacted NaCl(s) (see discussion in **§3.2.2** on quantification via measurement of integrated intensities and Rietveld refinement). Each CT powder pattern is indicative of a single crystallographically and magnetically distinct chloride ion in each crystal structure. In the case of NaCl:urea·$H_2O$, this is in accordance with the reported crystal structures,[87] whereas for LiCl:urea and CsCl:urea, this information is valuable for determining and refining their crystal structures (*vide infra*).

**3.2.2 Alkali Metal SSNMR Spectroscopy.** To investigate other quadrupolar NMR handles that might be useful for NMRX investigations and detection of educts and/or impurity phases, MAS NMR spectra of three alkali metal isotopes were acquired (**Figure 3**). In each of the alkali metal NMR spectra, the QI manifests differently, due to the distinct nuclear spin quantum numbers and relatively small nuclear quadrupole moments. Acquisition of static spectra for these nuclides is generally not necessary, since $^{23}$Na and $^{7}$Li tend to have very small CSAs (with powder patterns largely dominated by quadrupolar effects), $^{133}$Cs generally has small values of $C_Q$ and moderate CSAs (*vide infra*), and acquisition of their MAS spectra is facile.

The $^{23}$Na ($I$ = 3/2) MAS spectrum features two patterns: one broad pattern dominated by the effects of the SOQI and the sharp peak at $\delta_{iso}(^{23}Na)$ = 7 ppm, corresponding to the $Na^+$ cation site in the NaCl:urea·$H_2O$ and the NaCl educt, respectively. Since CT-selective π/2 pulses were used for acquisition of $^{23}$Na NMR spectra, the integrated peak intensities only provide approximate quantification of the amount of residual NaCl educt, since precise quantification would require significantly lower rf pulses.[114,116,117] Nonetheless, this fact, along with the use of long, calibrated recycle delays in $^{23}$Na SSNMR experiments, allows us to measure the integrated intensities of the sharp peaks and CT patterns, providing an estimation of 5 ± 2 w/w% residual NaCl. Rietveld refinement of the synchrotron PXRD pattern revealed the crystalline NaCl to amount to 0.6 wt%. There are several possible explanations that account for the discrepancy in the quantification of NaCl in the sample. The simplest explanation could be the samples for the NMR and synchrotron PXRD experiments are from different batches, and therefore have different amounts of residual NaCl.

By contrast, the $^{133}$Cs and $^{7}$Li MAS NMR spectra are different, with neither exhibiting a broad CT pattern. The $^{133}$Cs ($I$ = 7/2) MAS NMR spectrum of CsCl:urea has a single, sharp, CT peak flanked by series of sharp spinning sidebands, which is indicative of a single magnetically distinct $Cs^+$ site. Analysis reveals the presence of a substantial CSA (*i.e.*, a CS tensor with a span $\Omega$ = 235 ppm) and a very small $C_Q$, with the spinning sidebands largely arising from the influence of the former. Finally, the $^{7}$Li ($I$ = 3/2) SSNMR spectrum of LiCl:urea features a sharp CT peak that is flanked by many spinning sidebands arising from the +3/2 ↔ +1/2 and –1/2 ↔ –3/2 satellite transitions (STs), which are broadened by the first-order QI, and indicative of a single $Li^+$ cation in the asymmetric unit. There is no clear evidence of any $^{7}$Li resonances corresponding to residual LiCl(s) or LiCl(aq), which have values of $\delta_{iso}(^{7}Li)$ = –1.1 and 0.0 ppm,[118] respectively.

**3.2.3 $^{23}$Na SSNMR for Monitoring of the Decomposition of NaCl:urea·$H_2O$.** In preliminary $^{23}$Na MAS experiments at room temperature, changes observed in the $^{23}$Na SSNMR spectra over time suggest the gradual decomposition of NaCl:urea·$H_2O$ into its constituent components (NaCl, urea, and water). Furthermore, samples removed from the rotor revealed (i) a gradual change from a white crystalline powder to a grey paste and (ii) a narrow "bore hole" in the sample along the rotor axis, which is typically an indicator of water loss and/or sublimation. To monitor the decomposition of NaCl:urea·$H_2O$, a series of $^{23}$Na MAS NMR spectra were collected (**Figure 4**) using the following variable-temperature protocol: (i) A stationary sample was cooled to a nominal temperature of 5 ˚C, and then the rotor was spun up to 10 kHz, while maintaining this temperature (calibrations indicate an actual sample temperature of *ca*. 8 ˚C, due to frictional heating from MAS). (ii) An initial $^{23}$Na MAS NMR spectrum indicates that the sample is comprised largely of NaCl:urea·$H_2O$ with a small amount of NaCl(s). (iii) The sample was heated to 30 ˚C (calibrated temperature of *ca*. 37 ˚C under MAS), and $^{23}$Na MAS NMR spectra (recycle delay of 37.5 s and 16 scans) were acquired every 10 minutes over a period of three hours, leading to a total of 25 spectra. Over time, the $^{23}$Na MAS NMR spectra (**Figure 4A**) reveal the decomposition of NaCl:urea·$H_2O$ into the starting educts, as monitored quantitatively via measurement of integrated intensities of the patterns corresponding to NaCl:urea·$H_2O$ and NaCl.

A plot of the mol% of NaCl:urea·$H_2O$ (as approximated from the ratio of the integrated intensities of the powder patterns of NaCl:urea·$H_2O$ and NaCl) as a function of time (**Figure 4B**) reveals a sigmoidal shape that may be indicative of autocatalytic behaviour (*i.e.*, as NaCl:urea·$H_2O$ decomposes, the release of $H_2O$ serves to increase its rate of decomposition, until a steady state is achieved). If this is the case, the induction period has a duration of *ca*. 1200 seconds. Assuming the autocatalysis reaction follows an A + B → 2B reaction scheme, where A = NaCl:urea·$H_2O$ and B = $H_2O$, then a decomposition rate, *k*, which is assumed to be much larger than the reverse reaction, can be determined from the rate equation:

$$R = -\frac{\mathrm{d}[\mathrm{A}]}{\mathrm{d}t} = k[\mathrm{A}][\mathrm{B}] \quad [1]$$

If $[A]_0$ and $[B]_0$ represent the concentrations (or mol %) of A and B at the start of the reaction, then the total composition, $[C]_0$ is defined as $[C]_0 = [A]_0 + [B]_0 = [A] + [B]$ at any time during the reaction; therefore, $[A]_0 – [A] = [B] – [B]_0$. Thus, the rate equation can be rewritten as

$$-\frac{\mathrm{d}[\mathrm{A}]}{\mathrm{d}t} = k\{[\mathrm{A}][\mathrm{A}]_0 + [\mathrm{A}]_0[\mathrm{B}]_0 - [\mathrm{A}]^2\} \quad [2]$$

Eq. [2] can be integrated in with respect to [A] to yield:

$$[\mathrm{A}] = \frac{[\mathrm{A}]_0 + [\mathrm{B}]_0}{1 + \frac{[\mathrm{B}]_0}{[\mathrm{A}]_0}\{e^{([\mathrm{A}]_0+[\mathrm{B}]_0)kt}\}} = \frac{[\mathrm{C}]_0}{1 + \frac{[\mathrm{C}]_0 - [\mathrm{A}]_0}{[\mathrm{A}]_0}\{e^{([\mathrm{C}]_0)kt}\}} \quad [3]$$

Since the $[A]_f$, the final mol% of A is not zero, the integral constant F(*C*) is initially set equal to $[A]_f$:

$$[\mathrm{A}] = \frac{[\mathrm{C}]_0}{1 + \frac{[\mathrm{C}]_0 - [\mathrm{A}]_0}{[\mathrm{A}]_0}\{\mathrm{e}^{([\mathrm{C}]_0)\mathrm{kt}}\}} + \mathrm{F(C)} \quad [4]$$

A fit of Eq. [4] using a generalized reduced gradient nonlinear least squares method yields excellent agreement with experiment, with a decomposition rate constant of $k = 1.22 \times 10^{-3}$ $s^{-1}$ (**Figure 4**). Additional fits treating the system with a combination of autocatalytic (A + B → 2B) and direct (A → B) reactions with two distinct rates of $k_1$ and $k_2$, respectively, failed to yield fits of satisfactory quality. A more rigorous determination of the autocatalytic rate constants featuring both measurements and fits at multiple temperatures is beyond the scope of the current work, but is a clear point of interest for future studies of MCCs exhibiting either autocatalytic decomposition or growth.[119] We note that an analogous set of VT $^{35}$Cl SSNMR experiments were not conducted, due to their much longer experimental times and imperfect resolution of the two patterns corresponding to NaCl:urea·$H_2O$ and NaCl.

**3.2.4 SSNMR of Other Nuclides.** There are several other potentially useful NMR handles for which data is not reported in this work. First, $^{13}$C and $^{15}$N SSNMR data (collected with $^{1}$H-X CP/MAS experiments) are not reported due to the very long $T_1(^{1}\text{H})$ time constants, which would necessitate recycle delays on the order of 5.6 h, making these experiments impractical. This is in line with $T_1(^{1}\text{H})$ constants reported by Dybowski and co-workers for bulk urea.[120] Second, $^{13}$C and $^{15}$N SSNMR spectra were thought to be of limited value, since significant differences between chemical shifts among the various solid forms were not expected (later chemical shielding calculations revealed differences of less than 1 ppm). Third, direct excitation $^{13}$C and $^{15}$N experiments are similarly impractical, due to even longer $T_1(^{13}\text{C})$ and $T_1(^{15}\text{N})$ constants (this is especially problematic for $^{15}$N, with its low n.a. of ~0.37%). Finally, $^{14}$N SSNMR experiments were not attempted due to the large $C_Q$ value of 3.47 MHz in urea, lengthy experimental times even at very high fields, and the necessity for deuteration of samples to maximize $T_2^{\text{eff}}$ values.[121]

## 3.3 NMR-Guided Crystallography

The NMRX approach used in this work implements a combination of Rietveld refinement of high-quality synchrotron PXRD data, multinuclear SSNMR of quadrupolar nuclei, thermal analysis, and DFT calculations. Experimental synchrotron PXRD patterns were modeled and background corrected in the Materials Studio software package from BIOVIA, and indexed using the REFLEX module in Materials Studio (*N.B.*: Indexing the experimental synchrotron PXRD pattern is arguably the most challenging step in determining crystal structures of powdered samples because slight variations in the number of peaks and positions can lead to significant deviations). The X-cell algorithm was used to index the experimental synchrotron PXRD pattern and all potential solutions were subsequently subjected to a Pawley refinement to validate the indexing results. To generate an initial, feasible, structural model, the empty unit cells were populated with urea molecules, the appropriate ions, and water molecules in a stoichiometric ratio in accordance with data from multinuclear SSNMR and TGA/DSC analyses. Bond lengths and angles were corrected using the FORCITE molecular mechanics module in Materials Studio. These structural models were refined iteratively using REFLEX refinements and CASTEP geometry optimizations until a consistent solution was obtained. The final CASTEP model was validated using FOX,[111] which selected the space group with the highest possible symmetry for NaCl:urea·$H_2O$ ($I2$) and CsCl:urea ($P\bar{4}2_1/m$), and the non-standard setting $I2/c$ with β near 90° for the LiCl:urea (**Table S4**). The structural models were further optimized by Rietveld refinement in GSAS, initially using rigid body constraints on urea, followed by free refinement. Following this, hydrogen atoms were added to the structure in idealized positions. The resulting structures were subjected to final DFT-TS* geometry optimizations, enabling validation of the models by assessing the agreement between experimentally determined $^{35}$Cl EFG tensors with those obtained from the final refined structures (see **§3.4** for further details).

This NMRX approach was used to re-evaluate the structure of NaCl:urea·$H_2O$ and to obtain solutions and refinements of the structures of CsCl:urea and LiCl:urea, which are hitherto unknown (see **Figure S8** and **Table S4** for relevant crystallographic data). NaCl:urea·$H_2O$ crystallizes in the monoclinic *C*2 space group (No. 5, $Z$ = 4, $Z'$ = 1) with the following lattice parameters: $a$ = 6.4948 Å, $b$ = 5.2446 Å, $c$ = 17.3761 Å, β = 90.138º, $V$ = 591.874 $Å^3$ and is best described in the *I*2 setting where the β angle is close to 90°. The final Rietveld refinement (**Figure 5** and **Figure S8A**) gives an excellent fit of the synchrotron PXRD data ($R_p$ = 2.73 %, $R_{wp}$ = 3.18 %, $R_{F2}$ = 5.9 %), and stands in good agreement with the previously reported SCXRD structure.[87]

While the structure of NaCl:urea had been determined previously, CsCl:urea and LiCl:urea were unknown structures. The CsCl:urea MCC crystallizes in the tetragonal $P\bar{4}2_1/m$ space group (No. 113, $Z$ = 2, $Z'$ = 1), with unit cell parameters: $a$ = 5.796 Å, $c$ = 8.614 Å, $V$ = 289.375 $Å^3$. The final solution of the Rietveld refinement (**Figures 6** and **S8B**) shows excellent agreement with experimental synchrotron PXRD patterns ($R_p$ = 3.1 %, $R_{wp}$ = 2.1 %, $R_{F2}$ = 2.9 %). LiCl:urea (**Figures 7** and **S8C**) crystallizes in the monoclinic *C*2/*c* space group (No. 15, $Z$ = 16, $Z'$ = 2) with the following lattice parameters: $a$ = 14.4063 Å, $b$ = 8.867 Å, $c$ = 14.5611 Å, β = 91.219˚, $V$ = 1695.976 $Å^3$, in the non-standard setting *I*2/*c* , where the β angle is close to 90˚ and a relation to a tetragonal net is apparent.  The final Rietveld refinement solution shows good agreement with the experimental synchrotron PXRD patterns ($R_p$ = 1.9 %, $R_{wp}$ = 2.6 %, $R_{F2}$ = 4.4 %).

Interestingly, while both NaCl:urea·$H_2O$, and CsCl:urea MCCs exhibit only one alkali metal ion and one $Cl^-$ ion in the asymmetric unit, as demonstrated by their respective NMR spectra, LiCl:urea nominally contains two sets of crystallographically distinct $Li^+$ and $Cl^-$ species. A closer analysis suggests that the structure might best be described in the tetragonal space group $I4_1/a$, which would imply only one type of metal and halide ion. However, refinement of the crystal structure in this space group led to high temperature factors (the average experimental structure in this space group is shown in **Table S4**).

The crystal structure for NaCl:urea·$H_2O$ can be described as a layered structure consisting of alternating layers of MCl, urea, and water (**Figure 5**, inset). The $Na^+$ ions are in six-coordinate environments, coordinated to two $Cl^-$ ions, two oxygens from water molecules, and two oxygens from urea molecules. These layers are stacked along the crystallographic *c*-axis with the urea molecules directed at each other in a zigzag fashion. Similarly, the crystal structure for CsCl:urea features a layered structure consisting of CsCl and urea, with layers stacked along the crystallographic *c*-axis and the urea molecules oriented in a zigzag fashion (**Figure 6**, inset). However, the $Cs^+$ ions are in eight-coordinate environments, forming bonds to four $Cl^-$ ions and four oxygens from urea molecules. Lastly, the crystal structure for LiCl:urea adopts an arrangement analogous to analcime:[122] the $Li^+$ ions are in four-coordinate environments featuring two $Cl^-$ ions and two oxygen atoms from urea molecules, with the polyhedra alternating between edge-sharing and corner-sharing along the crystallographic *n*-axis (**Figure 7**, inset).

Of the three MCCs described herein, the crystal structure of LiCl:urea raises some questions. The best solution (*i.e.*, lowest energy) is consistent with the lower-symmetry subgroup, $I2/c$, of the tetragonal space group, $I4_1/a$. DFT-TS* geometry optimizations in the latter space group failed to converge. While the structural topology is consistent with a tetragonal net, the experimental observation and computational confirmation of a monoclinic space group with $\beta = 91.2^o$ implies a slight unit cell deformation of the tetragonal unit cell, perhaps due to molecular-level dynamics (see **Video S1)**. Since the structure with $\beta = 91.2^o$ has a symmetry-equivalent structure with $\beta = 88.8^o$, it is possible to refine a structure in the $I4_1/a$ space group, albeit with high temperature factors and the assumption that both structures are present in equal amounts. This is confirmed in part by the highly anisotropic thermal ellipsoids of the $Cl^{-}$ ions (**Figure 7**). From the diffraction data alone, it is not possible to ascertain the presence of dynamic or static disorder. However, dynamics of the urea molecules in the MCCs can significantly impact the SSNMR powder patterns of $^{35}Cl$, $^{14}N$, $^{2}H$, and quadrupolar alkali metal nuclides (for instance, see the $^{35}Cl$ SSNMR spectra in **Figure S9**).

To explore the possibility of dynamics influencing the observed $^{35}Cl$ tensor parameters for either LiCl:urea or CsCl:urea, we measured variable-temperature (VT) $^{35}Cl$ spectra at two temperatures, 298 K and 178 K, at 18.8 T (**Figure S9**). We elected to focus on LiCl:urea and CsCl:urea since the synchrotron PXRD and the Rietveld refinement suggested the possible presence of dynamics in these materials. We find that there are only minimal differences in the appearances of the spectra between these two temperatures, suggesting that molecular motions in this temperature range have little impact on the observed $^{35}Cl$ EFG and CS tensor parameters.

We also gave consideration to acquiring $^{6}Li$ SSNMR spectra in the hopes of resolving the two lithium shifts, which are not resolved in the $^{7}Li$ SSNMR spectrum, since the crystallographically unique $Li^{+}$ sites exist in almost identical environments. Since lithium has one of the smallest chemical shift ranges, it is very common to not be able to resolve such sites in $^{7}Li$ SSNMR spectra – in fact, our DFT calculations indicate that the lithium shifts differ by only 0.019 ppm, and would not be differentiated by either $^{7}Li$ or $^{6}Li$ SSNMR (chemical shift dispersion for $^{7}Li$ is 2.64× that of $^{6}Li$ at the same field). Furthermore, $^{6}Li$ SSNMR spectra are generally of higher resolution than those of $^{7}Li$, due to the lower quadrupole moment and reduced homonuclear dipolar couplings, but require much longer acquisition times (at natural abundance) due to low receptivity $R(^{6}Li)/R(^{7}Li) \approx 1/30$ and very long $T_1(^{6}Li)$ times.

### 3.4 DFT Calculations and EFG Tensor Orientations.

Quantum chemical computations are key in elucidating relationships between NMR interaction tensors and molecular-level structure.[40,57,82,123,124] Here, we consider a comparison of the experimentally-measured $^{35}Cl$ quadrupolar parameters with the $^{35}Cl$ EFG tensors and orientations obtained from DFT calculations. The $^{35}Cl$ EFG tensors are obtained from DFT calculations on the final refined structural models from the combination of Rietveld refinements and DFT-TS* geometry optimizations (see **§3.3**).[40] There is reasonably good agreement between the experimental and calculated $^{35}Cl$ EFG and CS tensors in all cases (**Table 1**).[57,58,82,83]

There are well-established relationships between $^{35}Cl$ EFG tensors of chloride ions and hydrogen bonding in organic solids (relationships between chlorine CS tensors and bonding are not straightforward, so further discussion is largely restricted to the EFG tensors).[55–57] A *hydrogen bond* is defined as having an H···Cl⁻ distance, $r$(H···Cl⁻), of 2.6 Å or less,[125] whereas a *short contact* is defined a hydrogen bond with $r$(H···Cl⁻) $\lesssim$ 2.2 Å. H···Cl⁻ short contacts exert the most influence on the $^{35}Cl$ EFG tensor parameters and orientations, with many organic HCl salts exhibiting values of $C_Q$ ranging from *ca*. 5 to 9 MHz in cases with one or two short contacts. The exceptions to this rule are H···Cl⁻ contacts involving $H_2O$ molecules, which present challenges for accurate calculations of $^{35}Cl$ EFG tensors.[58] These relationships have not been explored for MCCs featuring both chloride ions and alkali metal cations. Furthermore, for the three structural models herein, there are no short contacts (only between three and five hydrogen bonds) and relatively small magnitudes of $C_Q$ (**Table 2**). Nonetheless, it is worth investigating the orientations of the $^{35}Cl$ EFG tensors, to determine if they are constrained by local symmetry, local bonds, and/or crystallographic symmetry elements.

We now consider the relationships between the $^{35}$Cl EFG tensor orientations and local hydrogen-bonding environments of the chloride ions in the three MCCs (**Figure 8**). The chloride ions in NaCl:urea·$H_2O$ are defined by a single crystallographic position, are not located on crystallographic symmetry mirror planes or rotational axes (because of the low symmetry of the space group), and feature five hydrogen bonds involving urea (four) and water (one). Two monodentate urea molecules have the shortest H···$Cl^-$ bonds of $r$(H···$Cl^-$) = 2.383 and 2.475 Å with an H···$Cl^-$···H angle of ∠(H···Cl···H) = 129.3˚, and the single bidentate urea has $r$(H···$Cl^-$) = 2.486 and 2.503 Å with ∠(H···$Cl^-$···H) = 54.1˚. The largest component of the EFG tensor, $V_{33}$, is oriented approximately perpendicular to the bidentate H···$Cl^-$···H plane, with $V_{22}$ approximately in the plane and bisecting the ∠(H···$Cl^-$···H). $V_{33}$ is calculated to be negative (*i.e.*, $C_Q$ is positive, since $Q(^{35}\mathrm{Cl}) = -8.165\ \mathrm{fm}^2$), meaning the EFG diminishes along this axis moving away from the nucleus (*N.B.*: the sign of $C_Q$ cannot be determined from $^{35}$Cl SSNMR spectra, but can be calculated and used for interpreting relationships between structure and EFG tensors). The negative and positive signs of $V_{33}$ and $V_{22}$, respectively, as well as their orientations, are consistent with numerous predictions for similar H···$Cl^-$···H arrangements.[55–57,82]

In contrast to NaCl:urea·$H_2O$, the chloride ions in CsCl:urea are positioned in crystallographic mirror planes and on a two-fold rotational axis, with one set of H···$Cl^-$ hydrogen bonds from monodentate urea ligands reflected through the mirror plane ($r$(H···$Cl^-$) = 2.340), one set of bidentate H···$Cl^-$ hydrogen bonds in the mirror plane ($r$(H···$Cl^-$) = 2.494 Å and ∠(H···$Cl^-$···H) = 53.6°), and four identical Cs···$Cl^-$ contacts ($r$(H···$Cl^-$) = 3.514 Å). $V_{11}$ and $V_{22}$ are positioned in the crystallographic mirror plane, with $V_{33}$ oriented parallel to the mirror plane and perpendicular to the bidentate H···$Cl^-$···H plane. Opposite to the case of NaCl:urea·$H_2O$, $V_{33}$ is positive and $V_{11}$ bisects the ∠(H···$Cl^-$···H) plane.

LiCl:urea, in the monoclinic description, displays two crystallographically distinct chloride ions that are not associated with any symmetry operations in *I*2/c (neither local nor otherwise). Their local environments are almost identical, with three hydrogen bonds ranging from $r(\mathrm{H\cdots Cl^-}) = 2.354$ Å to $r(\mathrm{H\cdots Cl^-}) = 2.613$ Å, and two Li···Cl contacts (< 2.417 Å). In this system, which features the largest magnitudes of $C_Q$ in the MCl:urea·$n$$H_2O$ series, the values of $C_Q$ and $\eta_Q$ are calculated to be almost identical for the two sites ($C_Q$(Cl1) = 2.99 MHz, $\eta_Q$(Cl1) = 0.90, and $C_Q$(Cl2) = 2.94 MHz, $\eta_Q$(Cl2) = 0.90), with $V_{33}$ (which is negative) located approximately in an $\mathrm{H\cdots Cl^-\cdots H}$ plane formed by the shortest and longest hydrogen bonds, with an angle of $\angle(V_{33}\text{–}\mathrm{Cl^-\cdots H}) \approx 51^\circ$.

Several conclusions arise from the comparison of experimentally measured and computationally derived $^{35}$Cl EFG tensors, as well as consideration of their orientations: (i) even though the magnitudes of the QIs are small, the $^{35}$Cl EFG tensor parameters and orientations are highly dependent on local symmetry, hydrogen bonds (number, length, and moieties of origin), as well as crystallographic symmetry elements; and (ii) despite the absence of short $\mathrm{H\cdots Cl^-}$ contacts, the $^{35}$Cl EFG tensors orient themselves in a manner consistent with those previously described in the literature.[55,56,58,82,83]

## 4. Conclusions

Herein, we have demonstrated novel mechanochemical routes to synthesize three MCCs of the form MCl:urea·$n$$H_2O$ ($x$ = 0, 1), and conducted their structural characterization by integration of synchrotron PXRD, multinuclear SSNMR spectroscopy, Rietveld refinement, and plane-wave DFT calculations. The herein described MCCs can be prepared mechanochemically with high purity and great rapidity relative to more conventional cocrystallization through slow evaporation from water. The combination of PXRD and SSNMR analyses for the characterization of products of mechanochemical reactions proves to be extremely useful for rapid screening of products (as well as residual educts and/or impurities) and for the optimization of milling conditions (including milling time, milling media, and pH of the milling liquid). $^{35}$Cl SSNMR is well-suited for the structural characterization of these MCCs, since $^{35}$Cl EFG tensors are extremely sensitive to the smallest differences and/or changes in chloride ion environments, providing a powerful means of examining $H\cdots Cl^-$ bonding in organic solids. Alkali metal NMR proves to be useful for identifying the number of unique crystallographic sites and facile detection of educts and/or impurities; however, at the current time, it appears to be of limited use for aiding in structural refinements of the MCCs, since (i) their EFG tensors do not vary to the degree of those of $^{35}$Cl EFG tensors, largely due to the unresponsiveness of the alkali metal valence electrons to hydrogen bonding; and (ii) $^{7}$Li and $^{133}$Cs spectra of $Li^+$ and $Cs^+$ ions in organic solids generally have $C_Q$ values with very small magnitudes, which are challenging to calculate. Furthermore, it is possible that the dynamics of urea molecules in certain cases must be taken into account to make valid comparisons of experimental and calculated NMR interaction tensor data. $^{23}$Na MAS NMR was key for identifying residual NaCl educt, and proved to be beneficial for monitoring the degradation of NaCl:urea·$H_2O$ and extracting a decomposition rate constant of $k = 1.22 \times 10^{-3}$ s$^{-1}$. Finally, NMRX-guided Rietveld refinements, guided by comparison of $^{35}$Cl EFG tensors determined from experiment and computation, were used to resolve the structure of NaCl:urea·$H_2O$ and solve the novel structures of CsCl:urea and LiCl:urea. As a result, the insight garnered from this quadrupolar NMR-guided crystallography (*i.e.*, QNMRX) study of relatively simple MCCs with alkali metal cations and chloride anions will be useful for guiding future structural predictions and/or refinements of similar MCCs, and perhaps even new solid forms of pharmaceutical cocrystals featuring a wide range of pharmaceuticals, cations, anions, and other organic coformers.

**Declaration of Competing Interest**

The authors declare that they have no known competing financial interests or personal relationships that could have appeared to influence the work reported in this paper.

**Acknowledgements**

We thank the Florida State University, the National High Magnetic Field Laboratory, and the State of Florida, for support in the form of a startup grant. R.W.S. is grateful for research support from The Florida State University and the National High Magnetic Field Laboratory (NHMFL), which is funded by the National Science Foundation Cooperative Agreement (DMR-1644779, DMR-2128556), and by the State of Florida. A portion of this research used resources provided by the X-ray Crystallography Center at the FSU Department of Chemistry and Biochemistry (FSU075000XRAY). The XRD experiments used resources of the Advanced Photon Source Beamline 17-BM (XRD) at Argonne National Laboratory, which is an Office of Science User Facility operated for the U.S. Department of Energy (DOE) Office of Science and was supported by the U.S. DOE under contract no. DE-AC02-06CH11357. This research used resources of the Advanced Light Source, a U.S. DOE Office of Science User Facility under contract no. DE-AC02-05CH11231. E.B., and C.E.A.K. acknowledge funding from KULeuven (SIONA,C14/22/099). This work has received funding from the European Research Council (ERC) under grant agreement no. 834134 (WATUSO). E.B. acknowledges FWO for financial support (V401721N). NMRCoRe is supported by the Hercules Foundation (AKUL/13/21), by the Flemish Government as an international research infrastructure (I001321N), and by Department EWI via the Hermes Fund (AH.2016.134). Dr. Victor Terskikh is thanked for technical support and access to the 900 MHz NMR spectrometer, which was provided by the National Ultrahigh-Field NMR Facility for Solids (Ottawa, Canada), a national research facility funded by the Canada Foundation for Innovation, the Ontario Innovation Trust, Recherche Québec, the National Research Council of Canada, and Bruker BioSpin and is managed by the University of Ottawa (www.nmr900.ca). We thank the support of McGill University, NSERC Discovery Grant (RGPIN-2017-06467), Leverhulme International Professorship, and University of Birmingham.

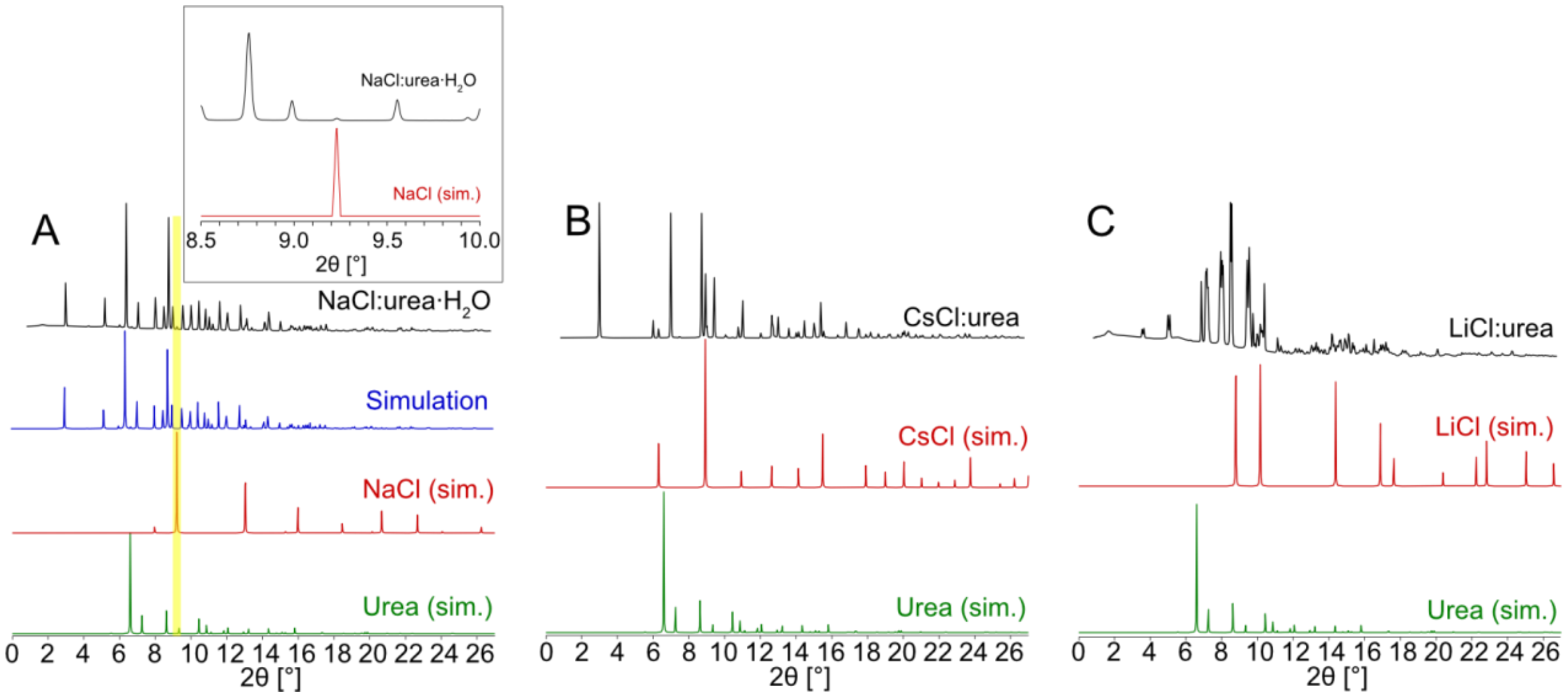

**Figure 1.** Experimental synchrotron PXRD patterns of (A) NaCl:urea·$H_2O$ (with the inset showing evidence of NaCl(s) impurity at 2θ = 9.22°), (B) CsCl:urea, and (C) LiCl:urea are shown in black, with simulated PXRD patterns of the educts, urea, and MCl salts, shown in green and red, respectively. In (A), a simulated PXRD pattern (in blue) based on the known structure of NaCl:urea·$H_2O$ is shown for comparison.[87] No evidence of residual educts are observed in the PXRD patterns of CsCl:urea and LiCl:urea.

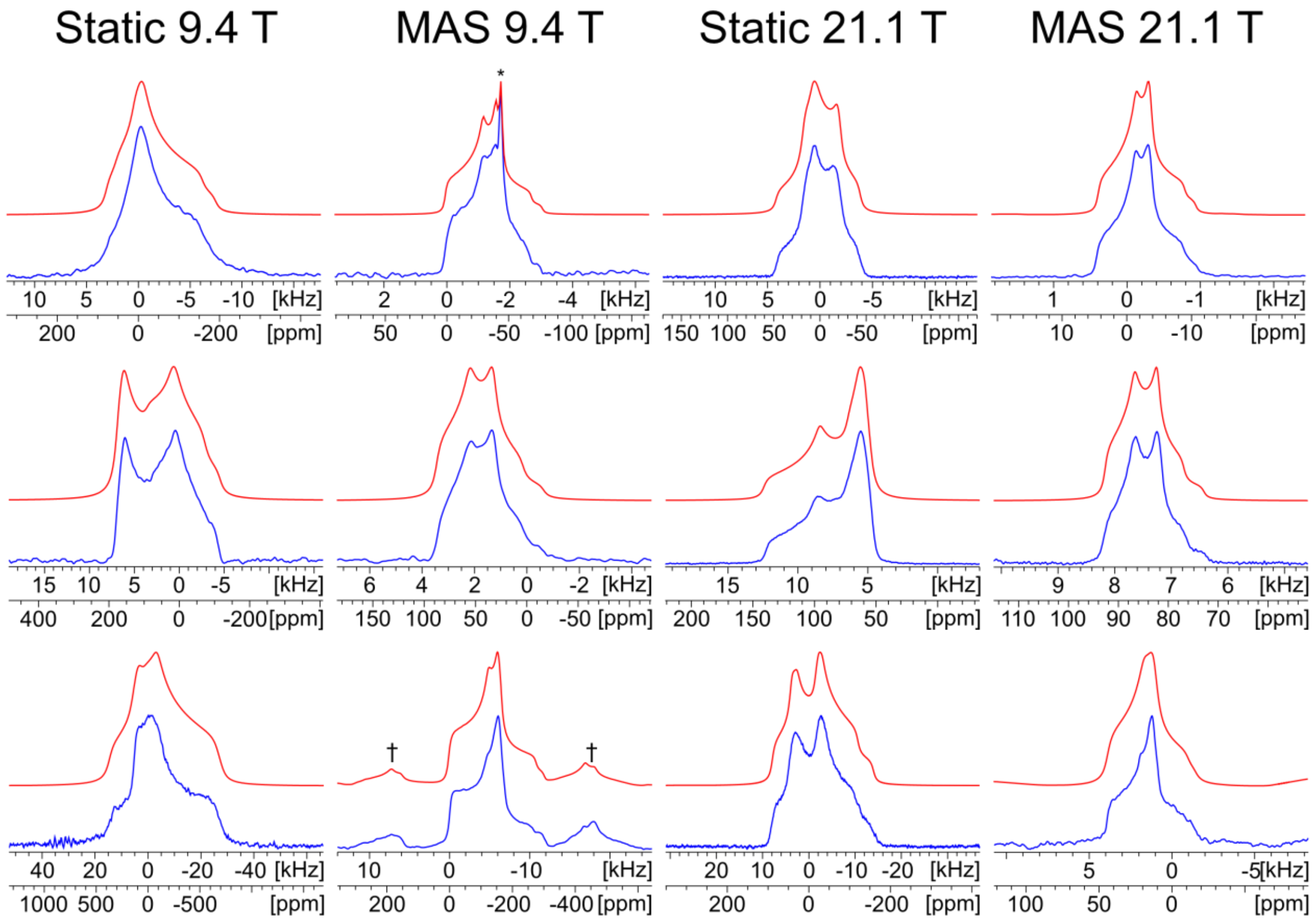

**Figure 2.** Experimental $^{35}Cl\{^{1}H\}$ SSNMR spectra for NaCl:urea·$H_2O$, CsCl:urea, and LiCl:urea (lower traces, blue) and corresponding analytical simulations (upper traces, red). Spectra were acquired at two fields ($B_0$ = 9.4 T and 21.1 T) under static and MAS conditions ($\nu_{rot}$ = 10 – 12 kHz). Spinning sidebands are indicated by †. In the MAS spectrum of NaCl:urea·$H_2O$ acquired at 9.4 T, a peak at $\delta_{iso}$ = –41.11 ppm (indicated by *) indicates a small amount of residual NaCl.

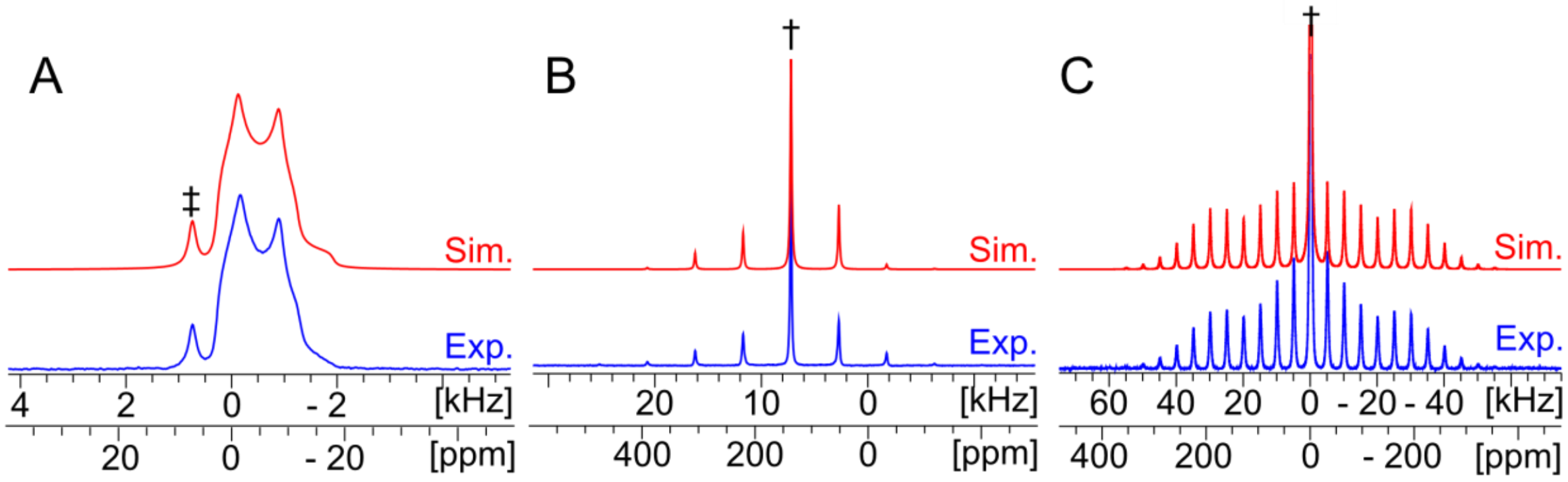

**Figure 3.** Experimental $^{23}Na$, $^{133}Cs$, and $^{7}Li$ SSNMR spectra of (A) NaCl:urea·$H_2O$, (B) CsCl:urea, and (C) LiCl:urea (lower traces, blue), with corresponding analytical simulations (upper traces, red). Spectra were acquired at 9.4 T under MAS conditions ($\nu_{rot}$ = 5 – 10 kHz). Isotropic peaks in (B) and (C) are labeled with a dagger (†). In the $^{23}Na$ spectrum in (A), a small amount of residual NaCl(s) is detected at $\delta_{iso}$ = 7 ppm (‡).

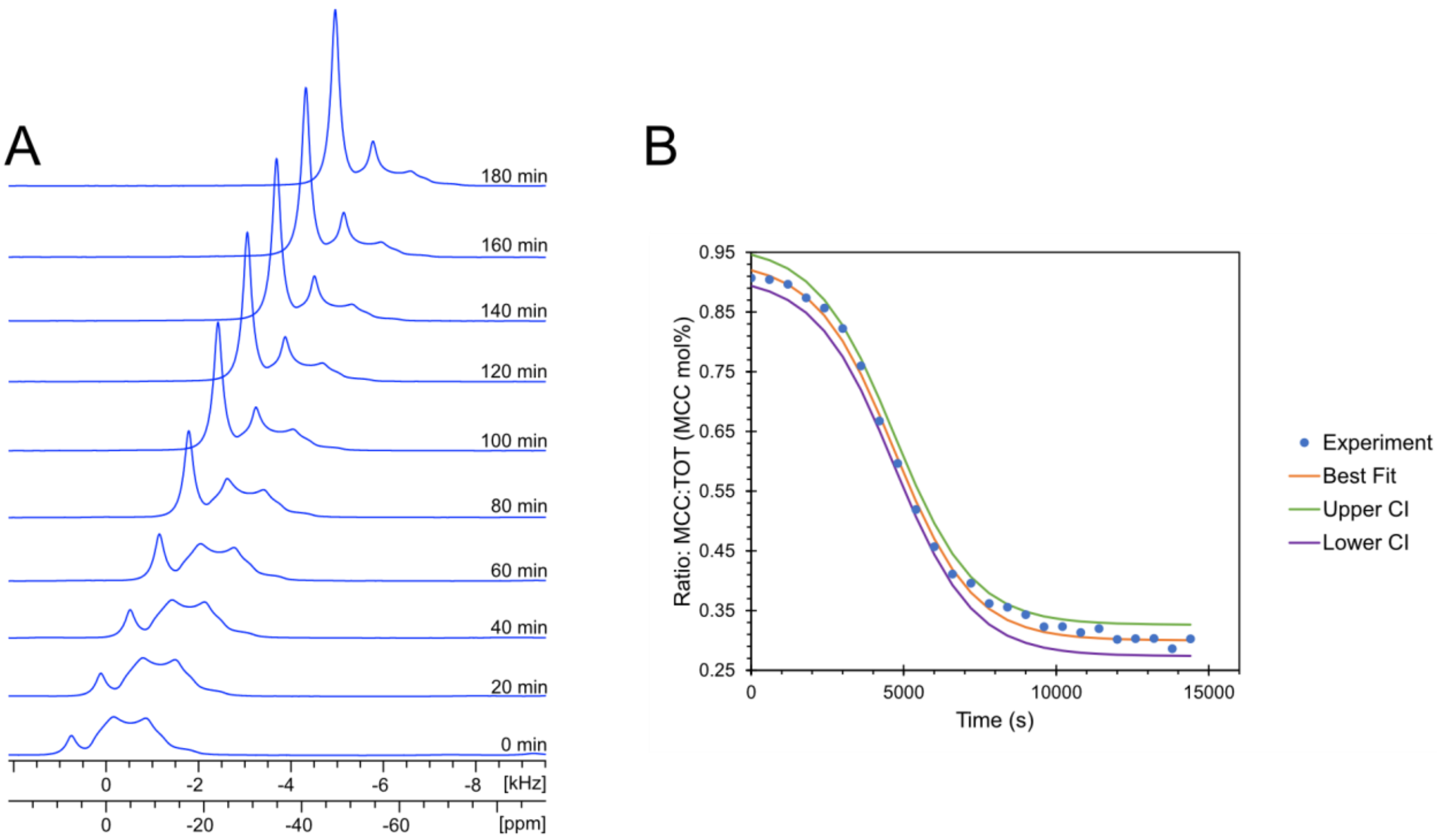

**Figure 4.** (A) Experimental $^{23}$Na MAS NMR spectra of NaCl:urea·$H_2O$ ($\nu_{rot}$ = 10 kHz) acquired at a fixed temperature of 30 °C (calibrated sample temperature at $\nu_{rot}$ = 10 kHz is *ca*. 37 °C) over three hours to monitor its degradation. (B) A plot of the mol% of NaCl:urea·$H_2O$ undergoing autocatalytic decomposition as a function of time (blue dots), along with a generalized reduced gradient nonlinear least squares fit using Eq. [4] (orange plot, see **§3.2.3** for details). Upper and lower confidence intervals are indicated by the green and purple lines.

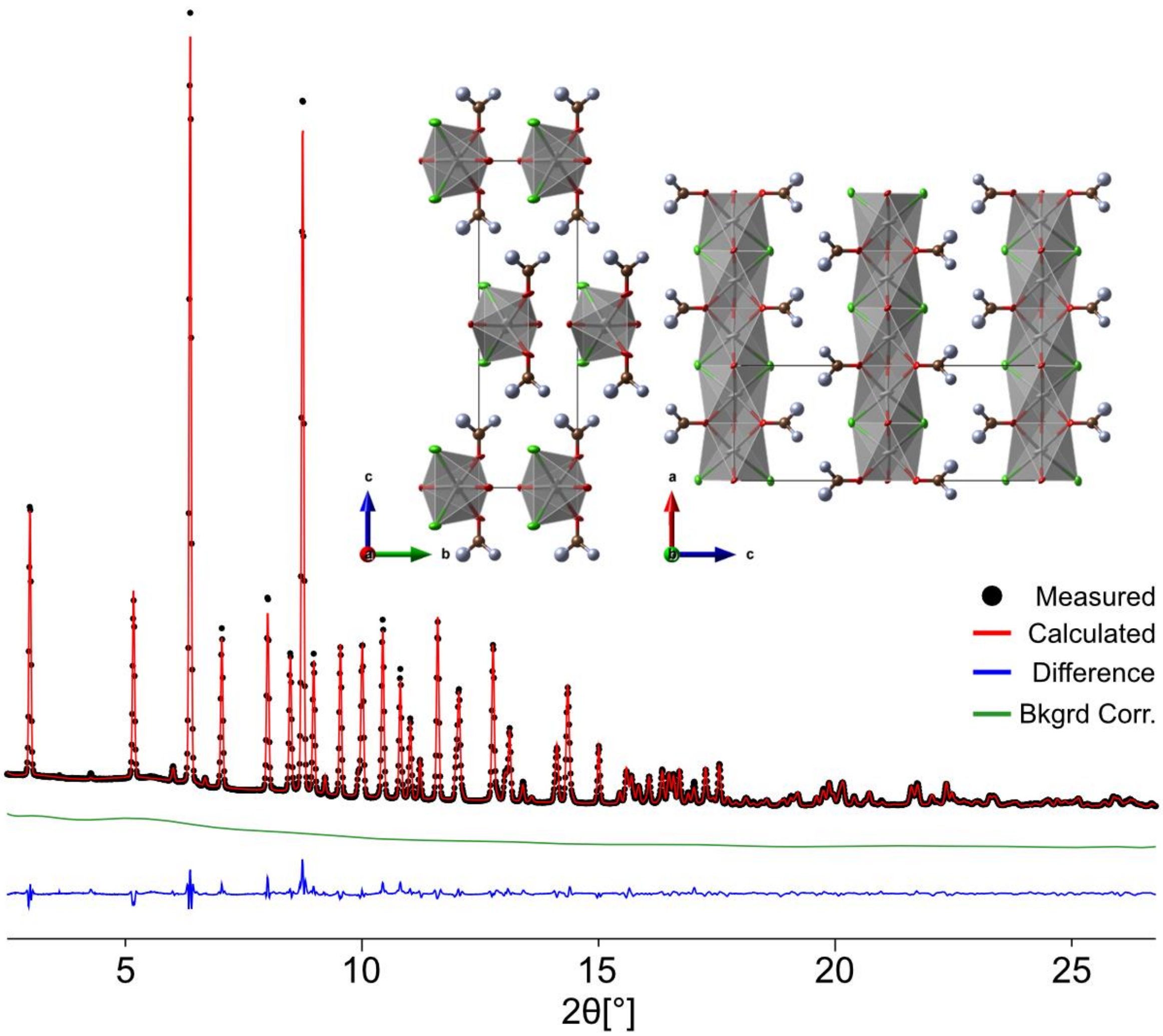

**Figure 5.** Rietveld plot for NaCl:urea·$H_2O$. Experimental data is shown in black and the calculated fit is shown in red. The difference plot is shown in blue and the background correction is shown in green. The inset (upper right) shows two views of the proposed crystal structure of NaCl:urea·$H_2O$. In the ball and stick model of the crystal structure, the different atoms are depicted as: carbon (brown), nitrogen (blue-grey), oxygen (red), chlorine (green), and sodium (grey). No hydrogen atoms are shown in these models.

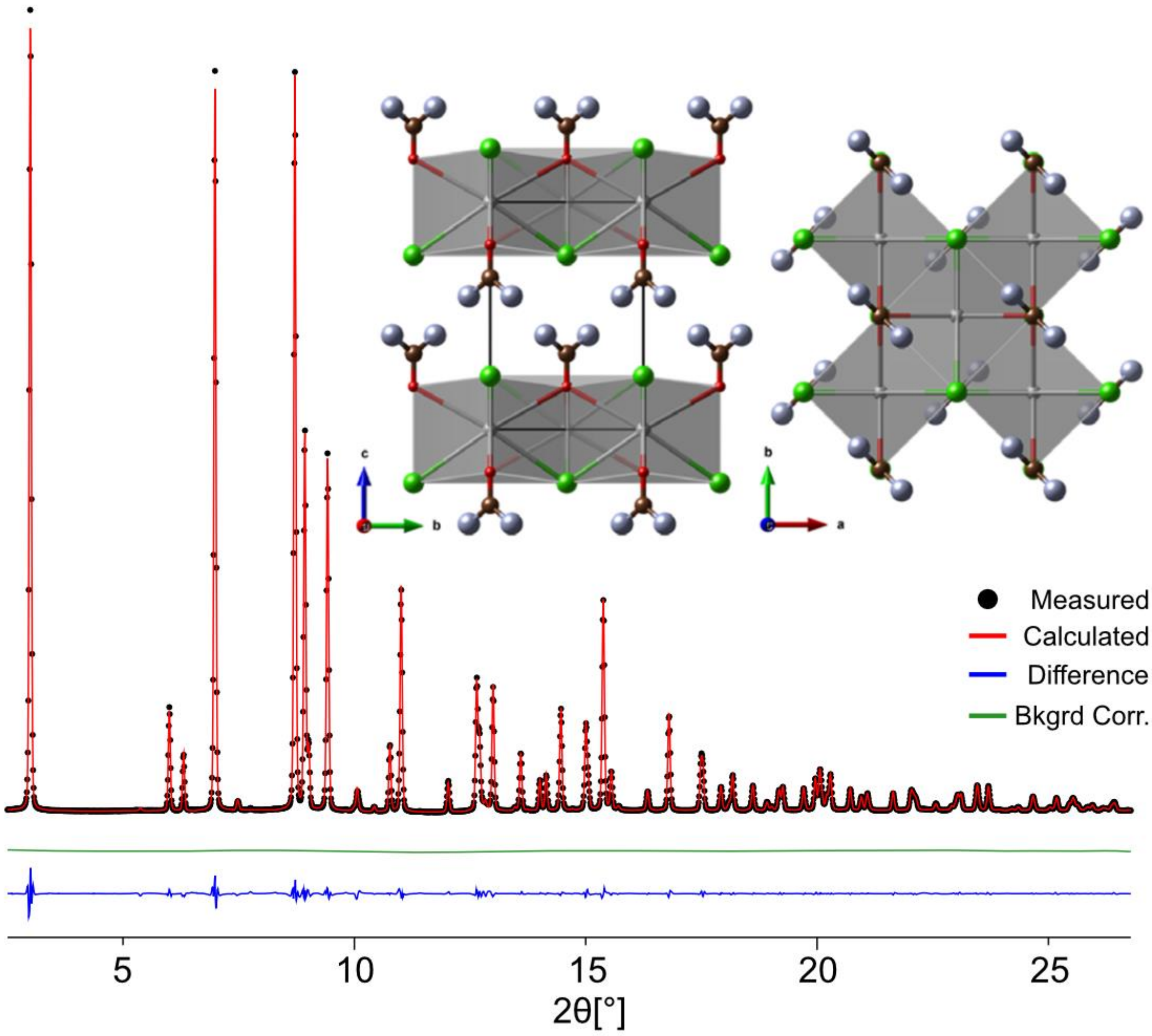

**Figure 6.** Rietveld plot for CsCl:urea. Experimental data is shown in black and the calculated fit is shown in red. The difference plot is shown in blue and the background correction is shown in green. The inset (upper right) shows two views of the proposed crystal structure of CsCl:urea. In the ball and stick model of the crystal structure, the different atoms are depicted as: carbon (brown), nitrogen (blue-grey), oxygen (red), chlorine (green), and cesium (grey). No hydrogen atoms are shown in these models.

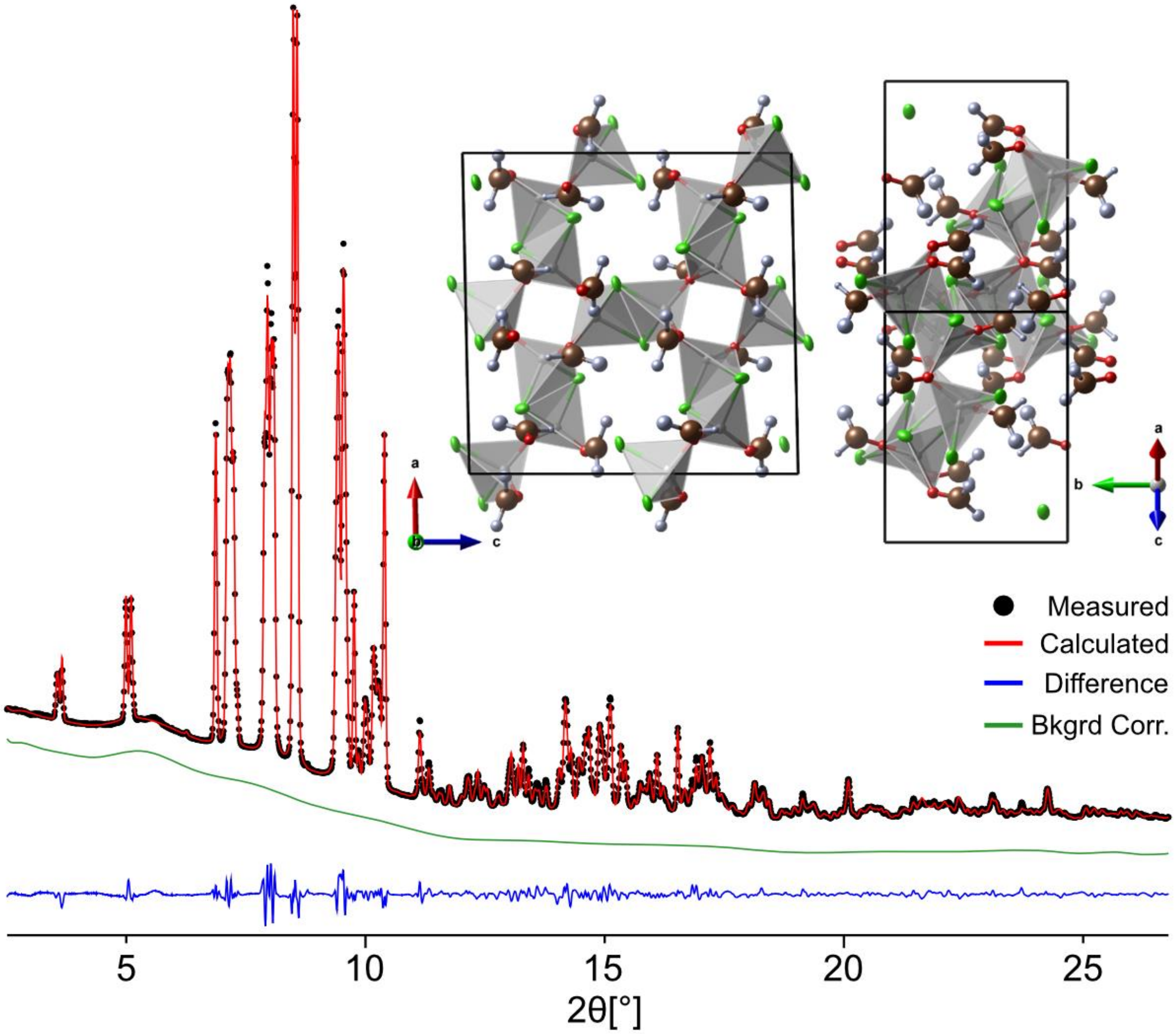

**Figure 7.** Rietveld refinement for LiCl:urea. Experimental data is shown in black and the calculated fit is shown in red. The difference plot is shown in blue and the background correction is shown in green. The inset (upper right) shows two views of the proposed crystal structure of LiCl:urea. In the ball and stick model of the crystal structure, the different atoms are depicted as: carbon (brown), nitrogen (blue-grey), oxygen (red), chlorine (green), and lithium (grey). No hydrogen atoms are shown in these models.

A

B

C

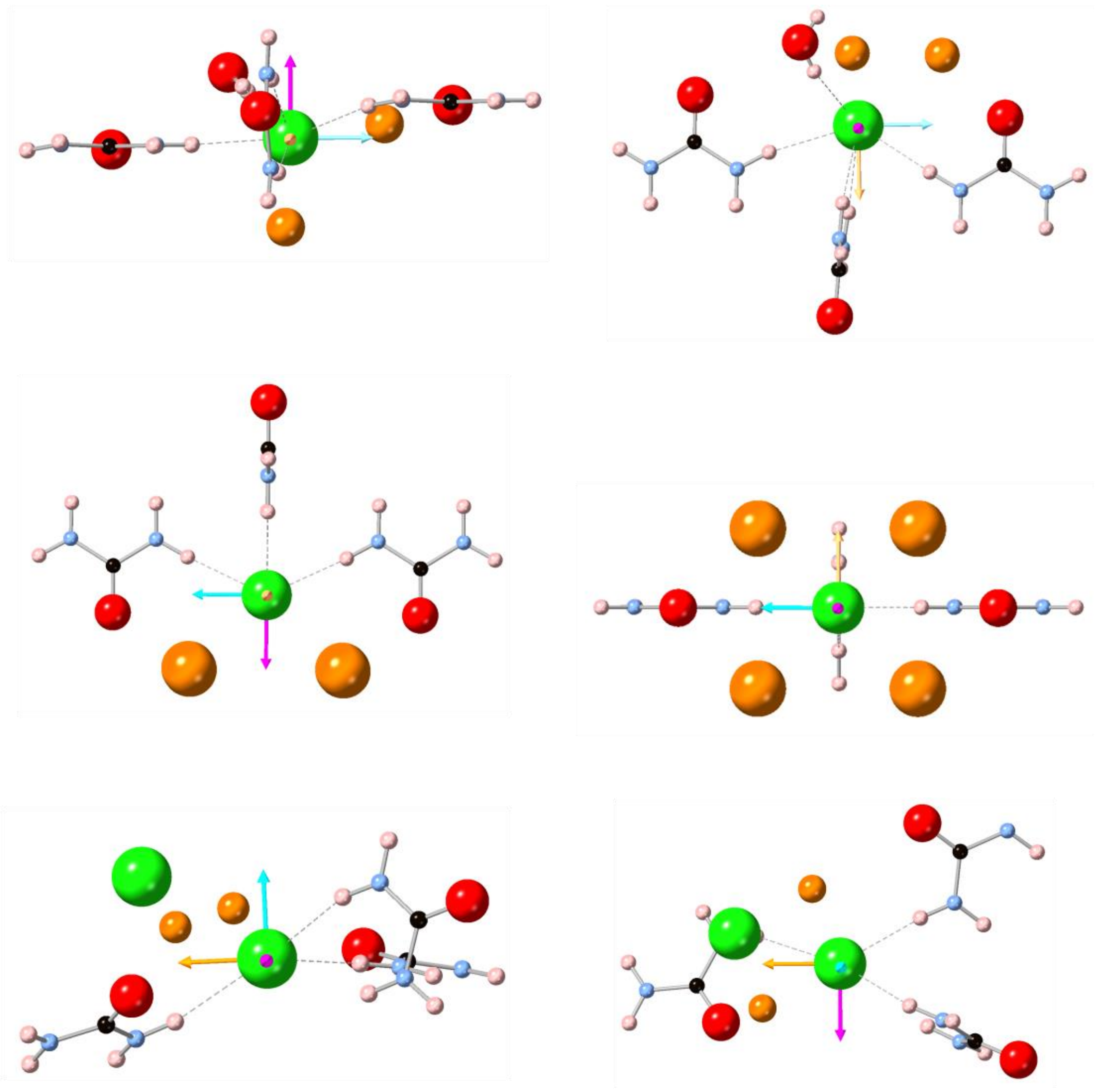

**Figure 8.** $^{35}$Cl EFG tensor orientations (two views for each system) for (A) NaCl:urea·$H_2O$, (B) CsCl:urea, and (C) LiCl:urea, obtained from calculations on structural models determined from Rietveld refinement and geometry optimization at the RPBE-TS* level. The H···$Cl^-$ hydrogen bonds (< 2.6 Å) are shown as black dashed lines. The orientations of the three principal components of the $^{35}$Cl EFG tensors are shown as magenta ($V_{11}$), yellow ($V_{22}$), and cyan ($V_{33}$).

**Table 1.** Experimental and calculated $^{7}$Li, $^{23}$Na, $^{133}$Cs and $^{35}$Cl EFG and CS tensors.†

| Material | Nucleus | | $C_Q$ [a] (MHz) | $\eta_Q$ [a] | $\delta_{iso}$ [b] (ppm) | $\Omega$ [b] (ppm) | $\kappa$ [b] | $\alpha$ [c] (°) | $\beta$ [c] (°) | $\gamma$ [c] (°) |
|---|---|---|---|---|---|---|---|---|---|---|
| NaCl:urea·$H_2O$ | $^{35}$Cl | Exp. | 1.37(5) | 0.73(8) | 4(2) | 75(10) | –0.35(8) | 65(5) | 81(2) | 4(3) |
| | $^{35}$Cl | Calc. [d] | 1.64 | 0.49 | –15 | 83 | –0.22 | 96 | 81 | 6 |
| NaCl:urea·$H_2O$ | $^{23}$Na | Exp. | 1.99(3) | 0.29(3) | 4.1(4) | – | – | – | – | – |
| | $^{23}$Na | Calc. | –2.08 | 0.43 | 2.2 | 10 | 0.50 | 30 | 19 | 47 |
| CsCl:urea | $^{35}$Cl | Exp. | 1.55(3) | 0.52(3) | 93(1) | 60(10) | –0.72(10) | 5(25) | 28(10) | 84(20) |
| | $^{35}$Cl | Calc. | –1.77 | 0.69 | 82 | 61 | –0.83 | 0 | 90 | 90 |
| CsCl:urea | $^{133}$Cs | Exp. | 0.106(5)† | 0.09(2) | 137(1) | 220(20) | –0.85(15) | – | – | – |
| | $^{133}$Cs | Calc. | –0.112 | 0.00 | 57.9 | 88 | –1.00 | 90 | 90 | 180 |
| LiCl:urea | $^{35}$Cl | Exp. | 2.63(2) | 0.85(3) | 4(3) | 77(20) | 0.50(10) | 75(15) | 20(10) | 160(20) |
| | $^{35}$Cl | Calc. (Cl1) | 2.99 | 0.90 | –10 | 78 | 0.80 | 80 | 86 | 143 |
| | $^{35}$Cl | Calc. (Cl2) | 2.94 | 0.90 | –9.4 | 79 | 0.63 | 80 | 84 | 125 |
| LiCl:urea | $^{7}$Li | Exp. | 0.093(4)† | 0.84(2) | 0.69(5) | – | – | – | – | – |
| | $^{7}$Li | Calc. (Li1) | –0.143 | 0.79 | 1.07 | 5 | –0.29 | 123 | 87 | 182 |
| | $^{7}$Li | Calc. (Li2) | –0.139 | 0.74 | 0.97 | 5 | –0.24 | 306 | 86 | 179 |

The experimental uncertainties in the last digit for each value are indicated in parentheses. Parameters indicated with "–" are not applicable or have little to no effect on the simulated $^{35}$Cl SSNMR,

† Indicates that only an upper limit can be estimated from simulations, due to the small magnitudes of the experimental parameters.

[a] The principal components of the EFG tensors are ranked as $|V_{33}| \geq |V_{22}| \geq |V_{11}|$. The quadrupolar coupling constant and asymmetry parameter are defined as $C_Q = eQV_{33}/h$, and $\eta_Q = (V_{11} - V_{22})/V_{33}$, respectively. The sign of $C_Q$ cannot be determined from the experimental $^{35}$Cl SSNMR spectra.

[b] The principal components of the chemical shift tensors are defined using the frequency-ordered convention such that $\delta_{11} \geq \delta_{22} \geq \delta_{33}$. The isotropic chemical shift, span, and skew are given by $\delta_{iso} = (\delta_{11} + \delta_{22} + \delta_{33})/3$, $\Omega = \delta_{11} - \delta_{33}$, and $\kappa = 3(\delta_{22} - \delta_{iso})/\Omega$, respectively.

[c] The Euler angles α, β, and γ define the relative orientation of the EFG and chemical shift tensors. Euler angles are reported using the *ZY′Z″* convention.[90–92]

[d] Theoretical EFG and CS tensor parameters were obtained from calculations on structures refined at the RPBE-TS* level.

**Table 2.** H···Cl⁻ hydrogen bonds (≲ 2.6 Å), contact angles, and calculated $^{35}$Cl EFG tensors.

| MCC | Hydrogen Bond Type [a] | H···Cl⁻ Distance [b] (Å) | X···Cl⁻ Distance [c] (Å) ‡ | X-H···Cl⁻ Angle [d] (˚) ‡ | $\delta_{iso}$ (ppm) | $C_Q$ (MHz) | $\eta_Q$ |
|---|---|---|---|---|---|---|---|
| NaCl:urea·$H_2O$ | NH···Cl⁻ | 2.383 | 3.332 | 169.9 | –15 | 1.64 | 0.49 |
| | NH···Cl⁻ | 2.475 | 3.42 | 168.5 | | | |
| | NH···Cl⁻ | 2.486 | 3.373 | 153.7 | | | |
| | NH···Cl⁻ | 2.503 | 3.385 | 152.7 | | | |
| | HOH···Cl⁻ | 2.240 | 3.133 | 163.0 | | | |
| CsCl:urea | NH···Cl⁻ | 2.340 | 3.297 | 176.5 | 82 | –1.77 | 0.69 |
| | NH···Cl⁻ | 2.340 | 3.297 | 176.5 | | | |
| | NH···Cl⁻ | 2.494 | 3.383 | 154.1 | | | |
| | NH···Cl⁻ | 2.494 | 3.383 | 154.1 | | | |
| LiCl:urea (Cl1) | NH···Cl⁻ | 2.366 | 3.322 | 173.9 | –10 | 2.94 | 0.90 |
| | NH···Cl⁻ | 2.419 | 3.35 | 163.1 | | | |
| | NH···Cl⁻ | 2.577 | 3.468 | 154.7 | | | |
| LiCl:urea (Cl2) | NH···Cl⁻ | 2.354 | 3.312 | 175.0 | –10 | 2.94 | 0.90 |
| | NH···Cl⁻ | 2.431 | 3.347 | 159.8 | | | |
| | NH···Cl⁻ | 2.613 | 3.496 | 153.2 | | | |

[a] The functional group involved in the H···Cl⁻ hydrogen bond (*e.g.*, NH···Cl⁻ and HOH···Cl⁻ signify hydrogen bonds with an amine functional group or water molecule and a chloride ion, respectively), as obtained from calculations on structures refined at the RPBE-TS* level.

[b] H···Cl– Hydrogen bonds (≤ 2.6 Å), as determined from crystal structures refined at the DFT-TS* level.

[c] Distance between the chloride ion and the hydrogen-bond donor atom.

[d] Angle between the hydrogen-bond donor atom, the hydrogen atom, and the chloride anion.

‡ X represents the atom bonded to hydrogen (*i.e.*, N or O).

*Electronic Supporting Information (ESI) for:*

# Rietveld Refinement and NMR Crystallographic Investigations of Multicomponent Crystals Containing Alkali Metal Chlorides and Urea

Cameron S. Vojvodin,[1,2] Sean T. Holmes,[1,2] Christine E. A. Kirschhock,[3] David A. Hirsh,[4] Igor Huskić,[5] Sanjaya Senanayake,[6] Luis Betancourt,[6] Weiqian Xu,[7] Eric Breynaert,[3] Tomislav Friščić,[5,9] and Robert W. Schurko[1,2]*

[1] Department of Chemistry & Biochemistry, Florida State University, Tallahassee, FL, 32306, USA
[2] National High Magnetic Field Laboratory, Tallahassee, FL, 32310, USA
[3] Centre for Surface Chemistry and Catalysis: KU Leuven, 3001 Leuven, Belgium
[4] Department of Chemistry & Biochemistry, University of Windsor, Windsor, Ontario N9B 3P4, Canada
[5] McGill University, Montreal, Quebec H3A 0B8, Canada
[6] Chemistry Division, Brookhaven National Laboratory, Upton, New York 11973, USA
[7] X-ray Science Division, Advanced Photon Source, Argonne National Laboratory, Argonne, Illinois 60439, USA
[8] NMR-Xray platform for Convergence Research (NMRCoRe), KU Leuven, 3001 Leuven, Belgium
[9] School of Chemistry, University of Birmingham, Edgbaston, Birmingham, B15 2TT, UK

*Author to whom correspondence should be addressed.
E-mail: rschurko@fsu.edu

**Table of Contents**

**Table S1.** Experimental conditions used in ball milling experiments for the syntheses of all MCCs described in this work.

| Material[a] | MCC Formation (Y/N) [d] | MCl (mol. eq.) [b] | Urea (mol. eq.) | Water (μL) | TEA (μL) [c] | Milling Freq. (Hz) | Milling Time (min) |
|---|---|---|---|---|---|---|---|
| LiCl:urea | Y | 1 | 1 | - | - | 30 | 40 |
| LiCl:urea·$H_2O$ | N | 1 | 1 | 18 | - | 30 | 40 |
| LiCl:urea·$H_2O$ | N | 1 | 1 | 18 | 10 | 30 | 40 |
| NaCl:urea | N | 1 | 1 | - | - | 30 | 40 |
| NaCl:urea·$H_2O$ | Y | 1 | 1 | 18† | - | 30 | 40 |
| NaCl:urea·$H_2O$ | Y | 1 | 1 | 18 | 10 | 30 | 40 |
| NaCl:urea·$H_2O$ | Y | 1 | 1 | 18 | 10 | 30 | 5 [d] |
| KCl:urea | N | 1 | 1 | - | - | 30 | 40 |
| KCl:urea·$H_2O$ | N | 1 | 1 | 18 | - | 30 | 40 |
| RbCl:urea | N | 1 | 1 | - | - | 30 | 40 |
| RbCl:urea·$H_2O$ | N | 1 | 1 | 18 | - | 30 | 40 |
| CsCl:urea | Y | 1 | 1 | - | - | 30 | 40 |
| CsCl:urea·$H_2O$ | N | 1 | 1 | 18 | - | 30 | 40 |
| CsCl:urea | Y | 1 | 1 | 18 | 10 | 30 | 40 |
| CsCl:urea | Y | 1 | 1 | 18 | - | 30 | 5 [e] |

[a] All reactions were scaled to yield ca. 200 mg of solid material.
[b] M = Li, Na, or Cs.
[c] Liquid assisted grinding (LAG) reactions involved the addition of μL quantities of triethylamine (TEA), as indicated.
[d] Y indicates an MCC was formed; N indicates no formation of an MCC (as determined by PXRD).
[e] Minimum time needed to prepare pure cocrystals by ball milling (BM), as indicated by PXRD.
† Stoichiometric water added to prepare hydrated MCCs.

**Table S2.** Experimental parameters for all SSNMR experiments in this work.

| Material | $B_0$ (T) | Nuc. | Expt | $\nu_{rot}$ (kHz) | Time Domain | Spectral Width (kHz) | Dwell Time (μs) | Acq. Time (ms) | Recycle Delay (s) | π/2[a] (μs) | Echo length (μs) | $^1$H dec (kHz) | Scans |
|---|---|---|---|---|---|---|---|---|---|---|---|---|---|
| LiCl:urea | 9.4 | $^7$Li | Hahn Echo | 5 | 4096 | 150 | 3.3 | 13.65 | 240 | 2.5 | 196 | 40 | 16 |
| | | $^{35}$Cl | Hahn Echo | 0 | 1024 | 200 | 2.5 | 2.56 | 2 | 2 | 213 | 40 | 12000 |
| | | $^{35}$Cl | Hahn Echo | 12 | 1024 | 100 | 5 | 5.12 | 2 | 2 | 164 | 40 | 4096 |
| | 21.1 | $^{35}$Cl | Quad Echo | 0 | 2048 | 100 | 5 | 10.24 | 15 | 3 | 50 | 35 | 3072 |
| | 21.1 | $^{35}$Cl | Bloch Decay | 5 | 4096 | 100 | 5 | 20.48 | 2 | 2 | - | 60 | 8192 |
| NaCl:urea·$H_2O$ | 9.4 | $^{23}$Na | Hahn Echo | 10 | 4096 | 50 | 10 | 40.96 | 25 | 1.25 | 98 | 40 | 32 |
| | 9.4 | $^{23}$Na | Bloch Decay | 10 | 4096 | 50 | 10 | 40.96 | 37.5 | 1.25 | - | - | 16 |
| | | $^{35}$Cl | Hahn Echo | 0 | 1024 | 50 | 10 | 10.24 | 4 | 2 | 205 | 40 | 5120 |
| | | $^{35}$Cl | Hahn Echo | 12 | 1024 | 50 | 10 | 10.24 | 4 | 2 | 80.33 | 40 | 512 |
| | 21.1 | $^{35}$Cl | Quad Echo | 0 | 2048 | 100 | 5 | 10.24 | 15 | 3 | 50 | 35 | 1024 |
| | 21.1 | $^{35}$Cl | Bloch Decay | 5 | 4096 | 100 | 5 | 20.48 | 5 | 2 | - | 60 | 512 |
| CsCl:urea | 9.4 | $^{133}$Cs | Hahn Echo | 4.5 | 4096 | 100 | 5 | 20.48 | 90 | 4 | 216 | 40 | 32 |
| | | $^{35}$Cl | Hahn Echo | 0 | 1024 | 100 | 5 | 5.12 | 3 | 2 | 125 | 40 | 6144 |
| | | $^{35}$Cl | Hahn Echo | 12 | 1024 | 50 | 10 | 10.24 | 3 | 2 | 80.33 | 40 | 1024 |
| | 21.1 | $^{35}$Cl | Quad Echo | 0 | 2048 | 100 | 5 | 10.24 | 20 | 3 | 46 | 35 | 1024 |
| | 21.1 | $^{35}$Cl | Bloch Decay | 5 | 4096 | 100 | 5 | 20.48 | 10 | 2 | - | 60 | 512 |

[a] calibrated central transition (CT) selective pulse

**Table S3.** Experimental parameters for all $^{35}Cl$ variable-temperature (VT) NMR QCPMG experiments conducted in this work at 18.8 T.

| Material | Temperature (K) | Time Domain | Spectral Width (kHz) | Dwell Time (μs) | Acq. Time (ms) | Recycle Delay (s) | $\pi/2^{a}$ (μs) | Echo length (μs) | Number of echoes | $^{1}H$ dec (kHz) | Scans |
|---|---|---|---|---|---|---|---|---|---|---|---|
| LiCl:urea | 298 | 8192 | 100 | 5 | 40.96 | 2 | 5 | 700 | 23 | 40 | 8192 |
| | 178 | 8192 | 100 | 5 | 40.96 | 2 | 5 | 700 | 23 | 40 | 256 |
| CsCl:urea | 298 | 8192 | 100 | 5 | 40.96 | 3 | 5 | 900 | 18 | 40 | 512 |
| | 178 | 8192 | 100 | 5 | 40.96 | 3 | 5 | 900 | 18 | 40 | 256 |

[a] calibrated central transition (CT) selective pulse

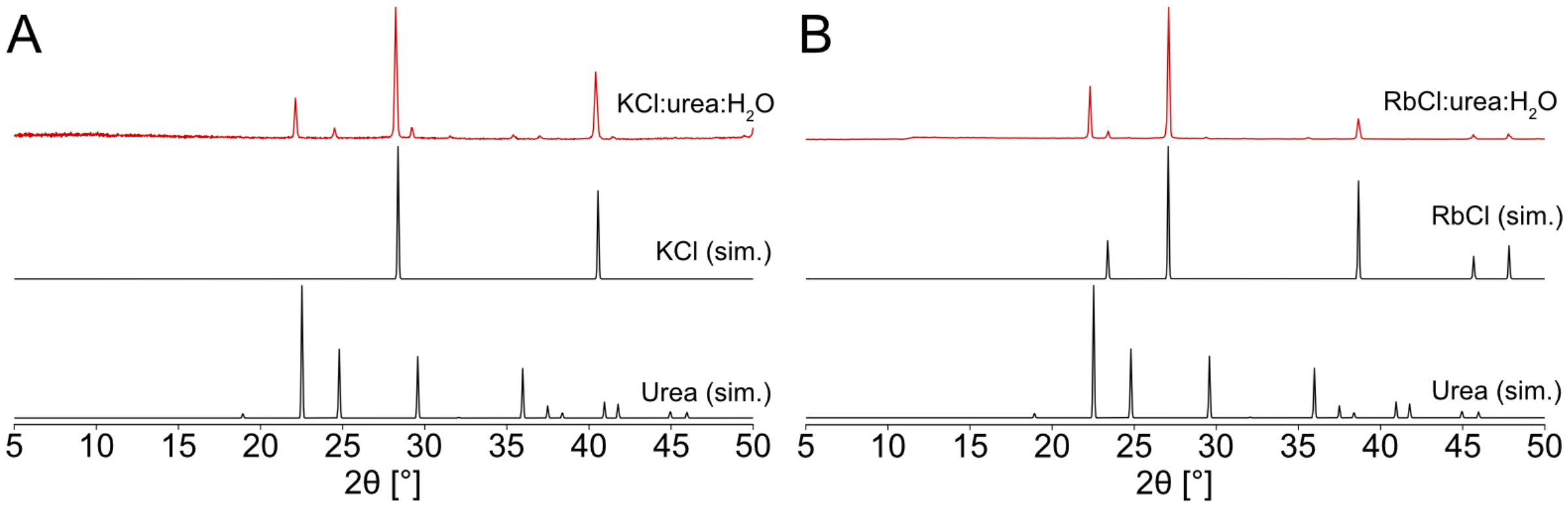

**Figure S1.** Experimental PXRD patterns (red) of LAG reaction mixtures of 1:1:1 ratios of MCl (M = K, Rb), urea, and $H_2O$ and simulated PXRD patterns (black) of urea and MCl salts. The PXRD patterns indicate that no MCCs are formed. PXRD patterns of NG reaction mixtures indicate identical results.

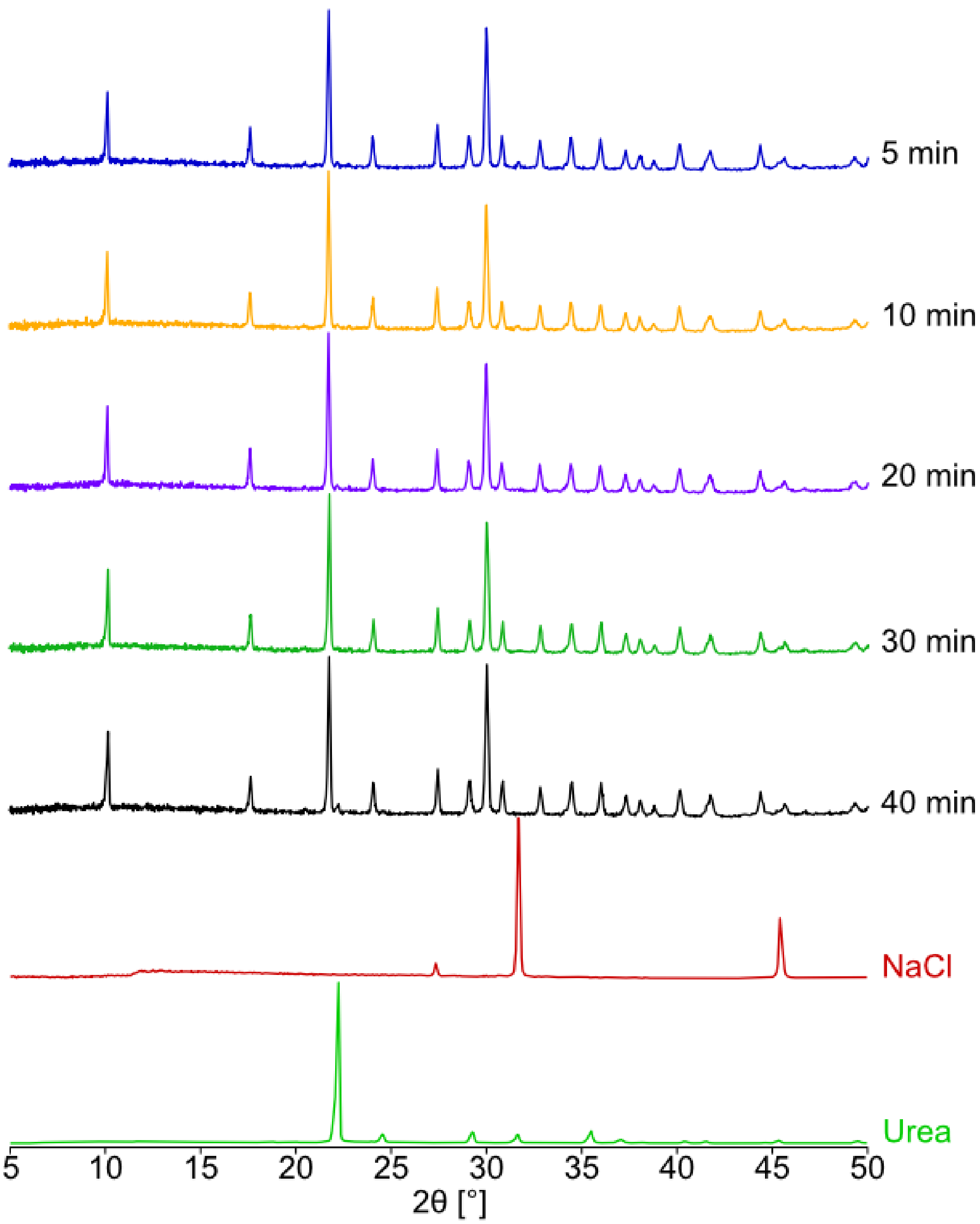

**Figure S2.** PXRD patterns of reaction mixtures of NaCl, urea, and $H_2O$ in a 1:1:1 ratio produced from LAG for 5, 10, 20, 30, and 40 minutes with a milling frequency of 30 Hz. The optimal milling time (*i.e.*, PXRD indicates the NaCl:Urea·$H_2O$ MCC as the sole product) is marked with an asterisk (*). Experimental PXRD patterns of NaCl and urea are shown for reference.

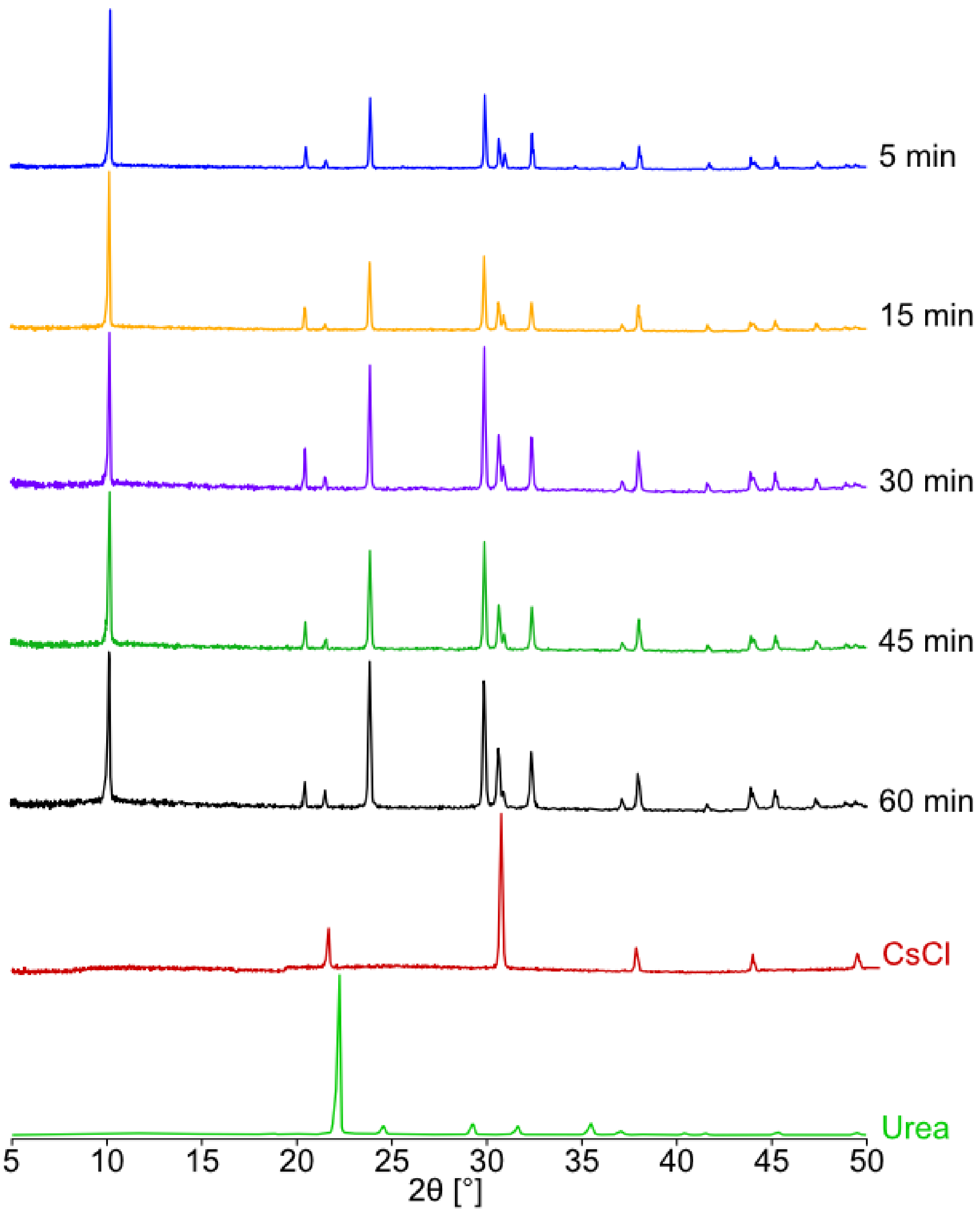

**Figure S3.** PXRD patterns for reaction mixtures of CsCl:Urea produced with NG for 5, 15, 30, 45, and 60 minutes with a milling frequency of 30 Hz. The optimal milling time is marked with an asterisk (*). Experimental PXRD patterns of CsCl and urea are shown for reference.

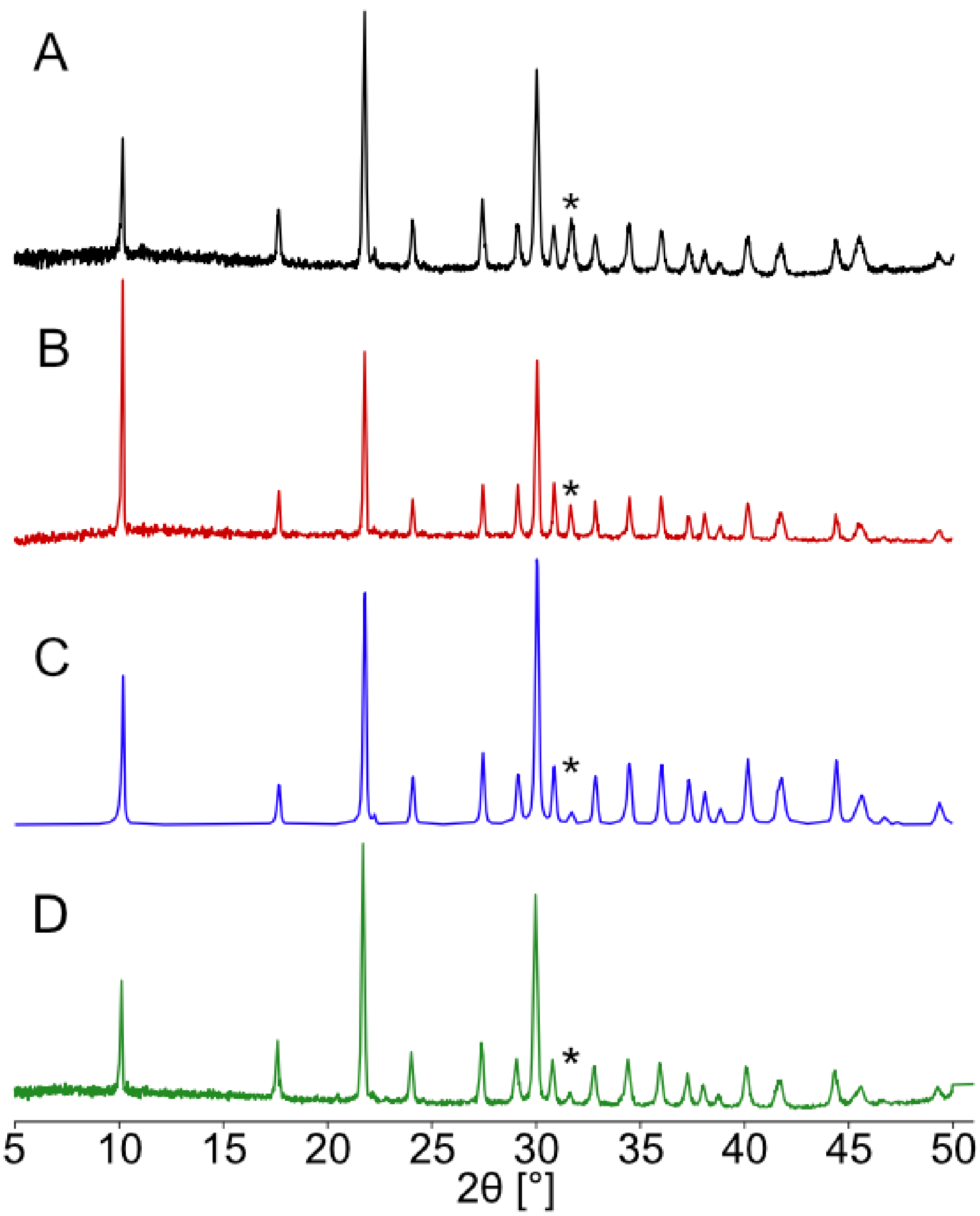

**Figure S4.** PXRD patterns for reaction mixtures of NaCl, urea and (A) 0.1 M $HNO_3$ in $H_2O$, (B) DI $H_2O$, (C) bottled water (Kirkland Signature$^{TM}$), and (D) $H_2O$ produced with LAG with the addition of 10 μL of triethylamine (TEA). In all cases, a small amount of residual NaCl (*) is present. However, the amount of NaCl:Urea·$H_2O$ product is diminished by the presence of carbonic acid (arising from atmospheric $CO_2$); the addition of a small amount of base to neutralize the acid increases its yield.

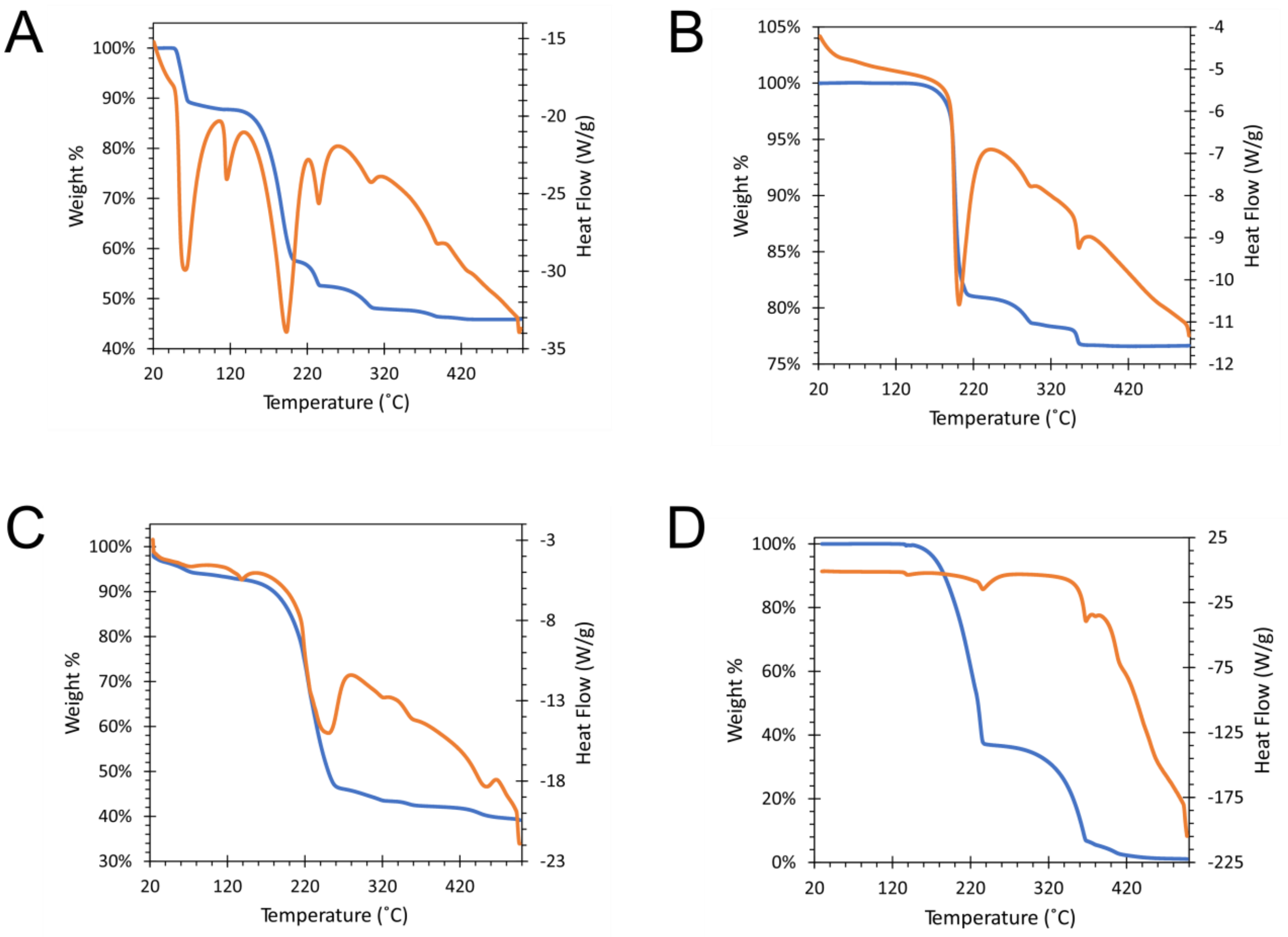

**Figure S5.** Thermogravimetric analysis (TGA; blue) and differential scanning calorimetry (DSC; orange) plots for (A) NaCl:urea·$H_2O$, (B) CsCl:urea, (C) LiCl:urea, and (D) urea. In panel A, a discontinuity in the TGA curve between 25 °C and 100 °C indicates the removal of stoichiometric water from the MCC. The weigh loss of *ca.* 13 % is consistent of one equivalence of water, confirming that NaCl:urea·$H_2O$ is a monohydrate. In panels B and C, this thermal event is asbsent indicating that both MCCs are anhydrous.

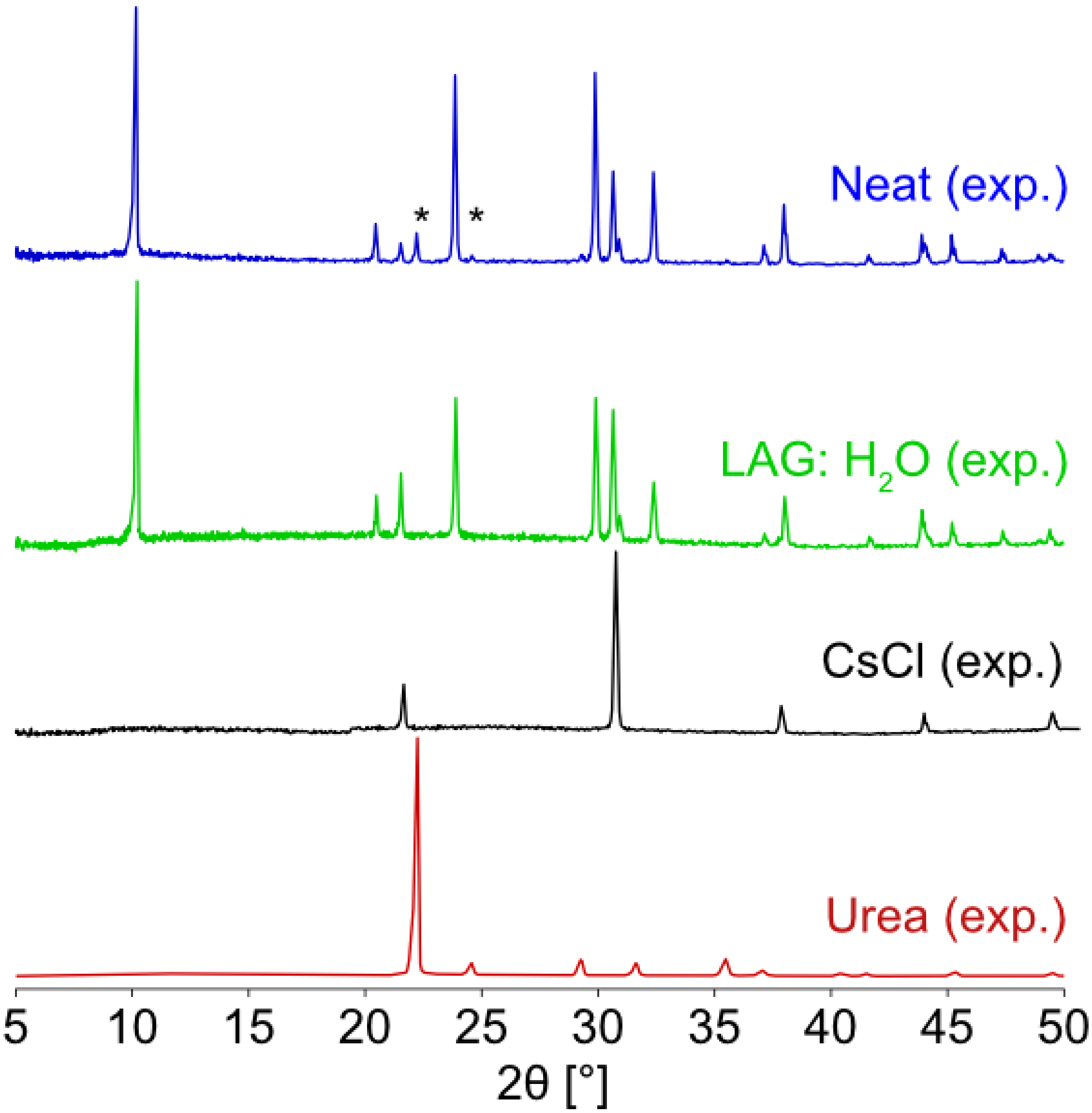

**Figure S6.** PXRD patterns of CsCl:urea prepared by NG (blue) and LAG (green) with water as the milling liquid. PXRD patterns of the urea (red) and CsCl (black) educts are shown above. The NG pattern (blue) shows that a small amount of residual urea, as indicated by the asterisk (*). Both preparations indicate the formation of CsCl:Urea with no leftover CsCl.

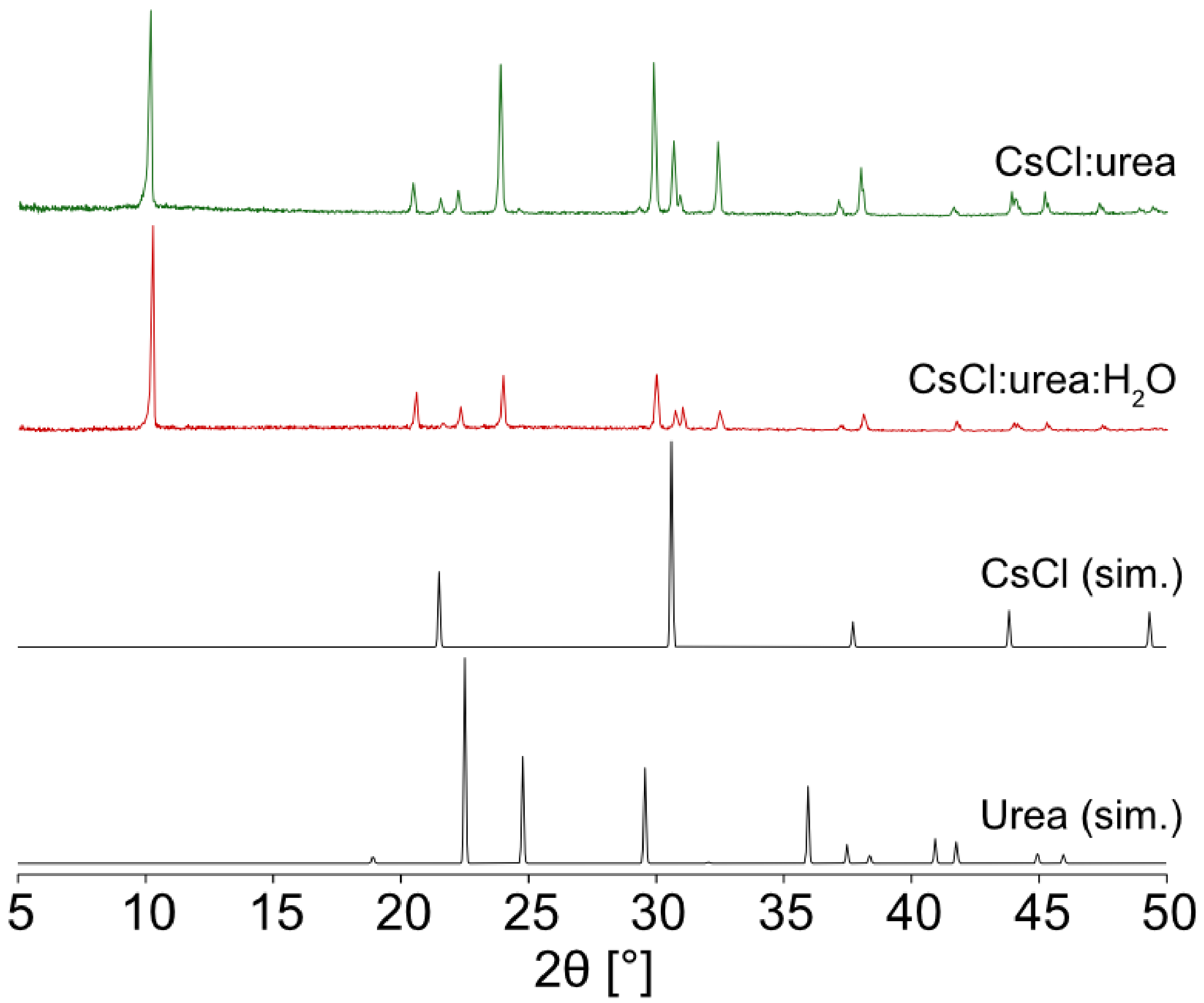

**Figure S7.** PXRD patterns of a 1:1 reaction mixture of CsCl:urea (green) prepared by NG and a 1:1:1 reaction mixture CsCl:urea:$H_2O$ (red) prepared by LAG, with water as the solvent. Simulated PXRD patterns of CsCl and urea are shown below (both in black). The resulting PXRD patterns for both reaction mixtures indicate the formation of the CsCl:urea MCC with no evidence of leftover educts.

**Table S4.** Summary of crystallographic information for the MCl:urea:$xH_2O$ MCCs.

| Material | NaCl:urea·$H_2O$ | NaCl:urea·$H_2O$ | CsCl:urea | LiCl:urea | LiCl:urea |
|---|---|---|---|---|---|
| CSD Code | NCUREA01 | - | - | - | - |
| Empirical Formula | $CH_6ClN_2NaO_2$ | $CH_6ClN_2NaO_2$ | $CH_4ClCsN_2O$ | $CH_4ClLiN_2O$ | $CH_4ClLiN_2O$ |
| Formula Weight (Da) | 136.47 | 130.47 | 252.40 | 98.42 | 98.42 |
| Crystal System | Monoclinic | Monoclinic | Tetragonal | Monoclinic | Tetragonal |
| Space Group | $I2$ | $I2$ | $P\bar{4}2_1/m$ | $I2/c$ | $I4_1/a$ |
| $a$ (Å) | 6.4845 | 6.454(7) | 5.796(6) | 14.4063(3) | 14.48366 |
| $b$ (Å) | 5.2362 | 5.212(5) | - | 8.867(12) | - |
| $c$ (Å) | 17.3497 | 17.272(17) | 8.614(9) | 14.5611(3) | 8.08666 |
| a (°) | - | - | - | - | - |
| β (°) | 90.152 | 90.131(2) | - | 91.219(2) | - |
| γ (°) | - | - | - | - | - |
| Cell volume (Å$^3$) | 589.092 | 581.0(18) | 289.3(9) | 1695.96(5) | 1696.39 |
| Calculated density (g cm$^{-3}$) | 1.5393 | 1.4916 | 2.6214 | 1.605 | 1.5415 |
| Cell formula units, $Z$ | 4 | 4 | 2 | 16 | 16 |
| Cell asymmetric units, $Z'$ | 1 | 1 | 1 | 2 | 2 |
| $R_p$ [a] | - | 0.032 | 0.032 | 0.019 | n/a |
| $R_{wp}$ [b] | - | 0.021 | 0.021 | 0.026 | n/a |
| $R_{exp}$ [c] | - | 0.014 | 0.014 | 0.022 | n/a |
| $R(F^2)$ [d] | - | 0.0290 | 0.0290 | 0.04354 | n/a |
| GoF [e] | - | 1.81 | 1.6 | 1.24 | n/a |

[a] Residual of least-squares refinement
[b] Weighted $R$-factor
[c] Expected $R$-factor
[d] $R$-factor based on $F^2$
[e] Goodness of fit = comparison of the $R_{wp}$ and $R_{exp}$

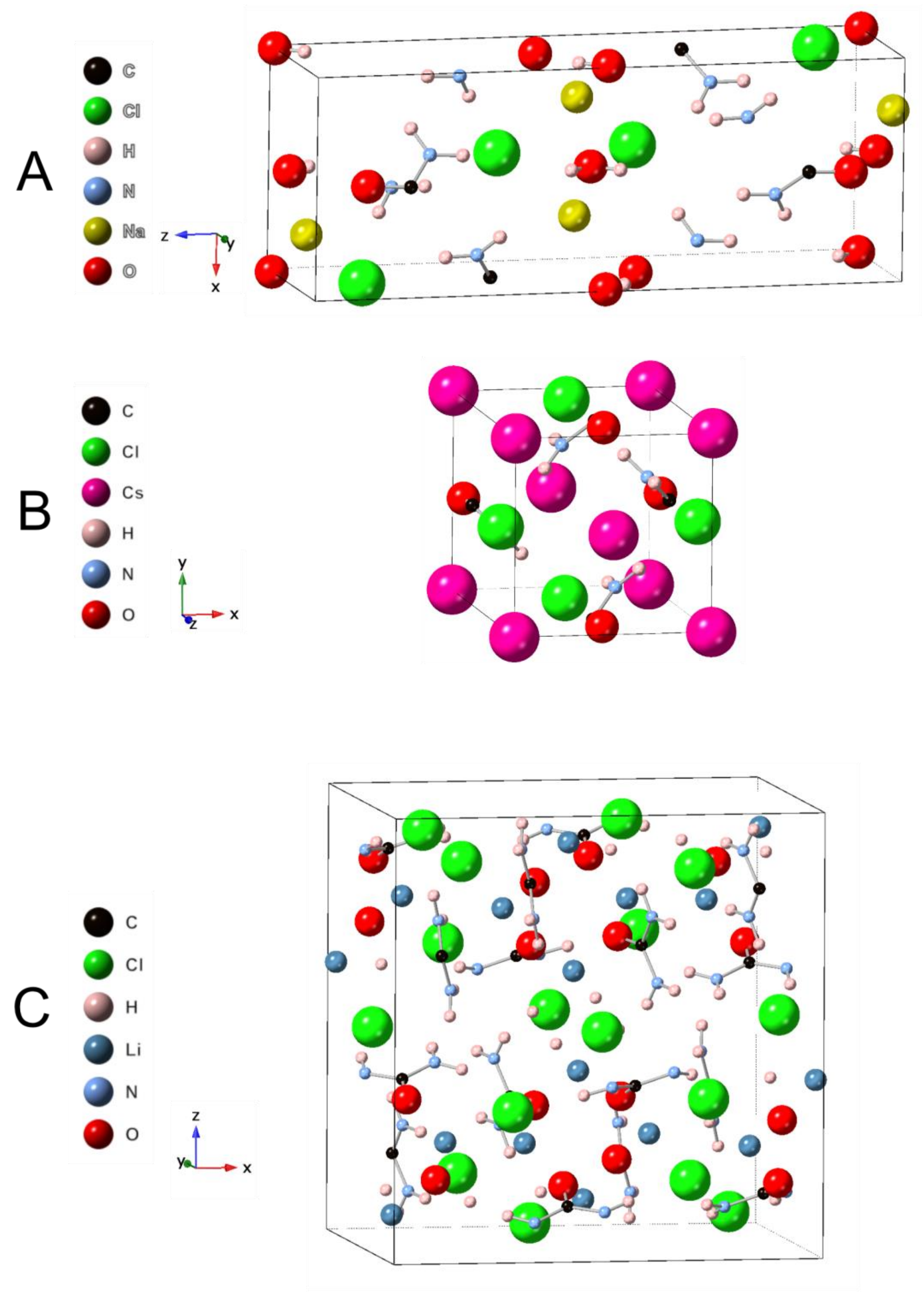

**Figure S8.** Images of the geometry optimized unit cells solved by NMRX-guided Rietveld refinements for (A) NaCl:urea·$H_2O$, (B) CsCl:urea, and (C) LiCl:urea.

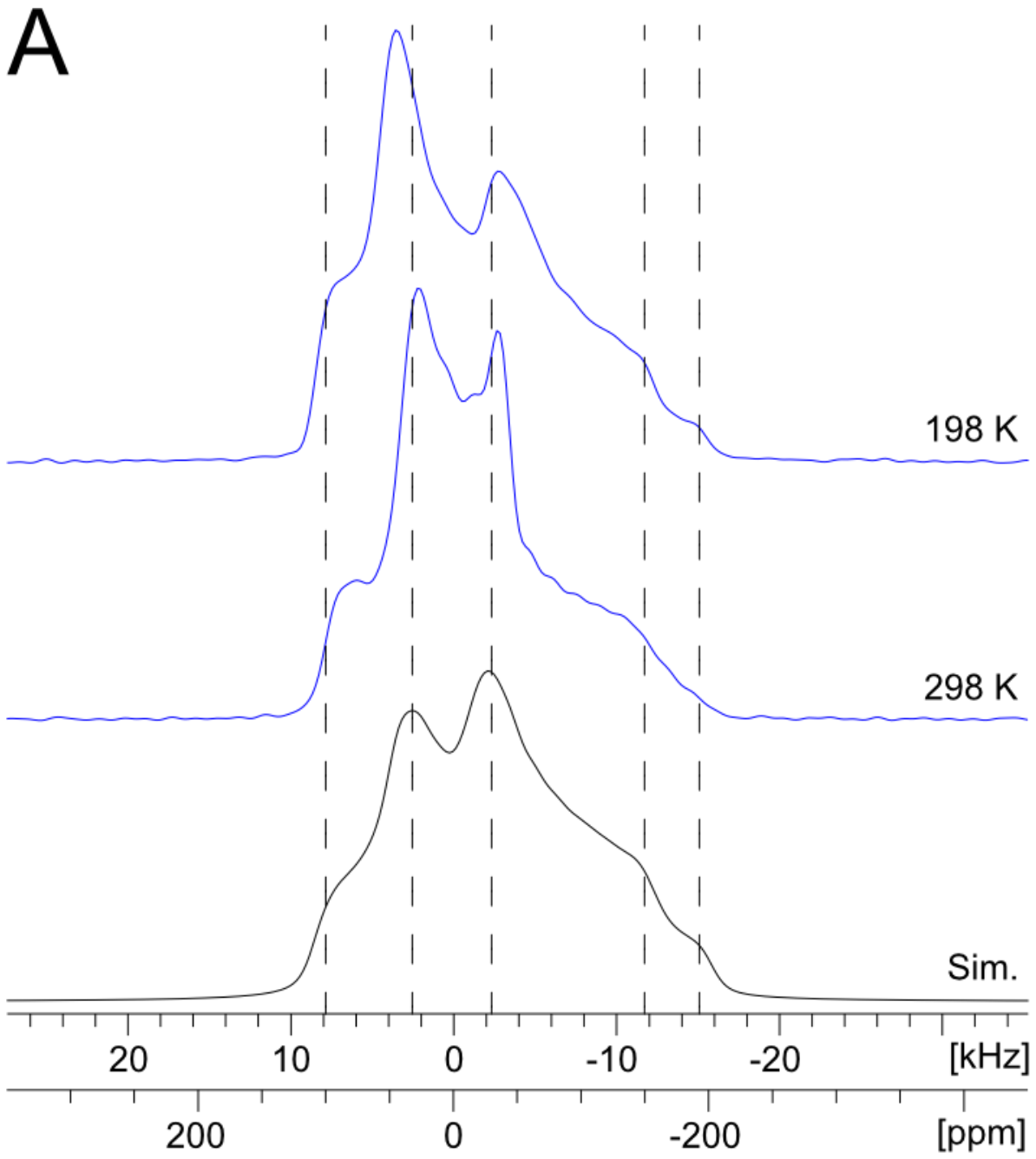

**Figure S9.** Experimental $^{35}Cl\{^{1}H\}$ CPMG VT-NMR spectra (upper traces, blue) acquired at $B_0$ = 18.8 T at 298 K and 178 K for (A) LiCl:urea and (B) CsCl:urea with corresponding analytical simulations (lower traces, black). The black dashed lines indicate key discontinuities of the simulated pattern. Both spectra (A) and (B) show minimal change in the $^{35}Cl$ EFG tensor parameters with decreasing temperature.